\documentclass[aps,floatfix,onecolumn,preprint]{revtex4-1}

\usepackage{dcolumn}
\usepackage{bm}
\usepackage{amsmath}
\usepackage{isomath}
\usepackage{psfrag}
\usepackage{subfigure}
\usepackage{times}
\usepackage{mathptmx}
\usepackage{verbatim}
\usepackage{graphicx}
\usepackage{color}
\usepackage{soul}
\usepackage{placeins}
\usepackage{amssymb}

\def\be{\begin{equation}}
\def\ee{\end{equation}}
\def\bes{\begin{equation*}}
\def\ees{\end{equation*}}
\def\bea{\begin{eqnarray}}
\def\eea{\end{eqnarray}}
\def\beas{\begin{eqnarray*}}
\def\eeas{\end{eqnarray*}}
\def\bal#1\eal{\begin{align}#1\end{align}}
\def\bals#1\eals{\begin{align*}#1\end{align*}}
\newcommand{\bra}[1]{\langle #1|}
\newcommand{\ket}[1]{|#1\rangle}

\renewcommand{\vec}{\vectorsym}
\newcommand{\del}{\partial}

\graphicspath{{pictures/}}

\begin{document} 

\title{Experimental evidence for Wigner's tunneling time}

\author{Nicolas Camus,$^{\ast}$, Enderalp Yakaboylu,$^{\ast \dag}$ Lutz Fechner, Michael Klaiber, \\ Martin Laux, Yonghao Mi, Karen Z. Hatsagortsyan, \\
 Thomas Pfeifer, Christoph H. Keitel, and Robert Moshammer}

\affiliation{Max-Planck-Institut f\"{u}r Kernphysik, Saupfercheckweg 1, 69117 Heidelberg, Germany\\
\\
\normalsize{$^\dag$ present address: IST Austria (Institute of Science and Technology Austria), }\\
\normalsize{ Am Campus 1, 3400 Klosterneuburg, Austria }\\
\\
\normalsize{$^\ast$These authors contributed equally to the work:}\\ 
\normalsize{N. C. for the experimental and E. Y. for the theory side.}}


\begin{abstract}
\textbf{ Tunneling of a particle through a potential barrier - i.e. its presence in a classically forbidden region - remains one of the most remarkable quantum phenomena. Owing to advances in laser technology, electric fields comparable to those electrons experience in atoms are readily generated \cite{Brabec2000} and open opportunities to dynamically investigate the process of electron tunneling through the potential barrier formed by the superposition of both laser and atomic fields \cite{Keldysh1965}. Attosecond-time \cite{Eckle2008} and angstrom-space resolution of the strong laser-field technique \cite{Krausz2009} allow to address fundamental questions related to tunneling, which are still open and debated\cite{Pfeiffer2012,Landsman2014o,Yakaboylu2014,Ni2016,Torlina2015}: Which time is spent under the barrier and what momentum is picked up by the particle in the meantime? In this combined experimental and theoretical study we demonstrate that for strong-field ionization the leading quantum mechanical Wigner treatment for the time resolved description of tunneling is valid. We achieve a high sensitivity on the tunneling barrier and unambiguously isolate its effects by performing a differential study of two systems with almost identical tunneling geometry. Moreover, working with a low frequency laser, we essentially limit the non-adiabaticity of the process as a major source of uncertainty.
The agreement between experiment and theory implies two substantial corrections with respect to the widely employed quasiclassical treatment: In addition to a non-vanishing longitudinal momentum along the laser field-direction we provide clear evidence for a non-zero tunneling time delay. 
This addresses also the fundamental question how the transition occurs from the tunnel barrier to free space classical evolution of the ejected electron. Given that we find tunneling times on the order of tens of attoseconds, our work has relevant applications for attosecond spectroscopy with nowadays resolution on this order because common techniques such as high-harmonic generation \cite{Corkum2007,Calegari2016} or laser-induced electron diffraction \cite{Meckel2008,Blaga2012} rely also on tunnel ionization.}

\end{abstract}

\maketitle

\begin{figure}
  \centering
\includegraphics[width=0.84\linewidth]{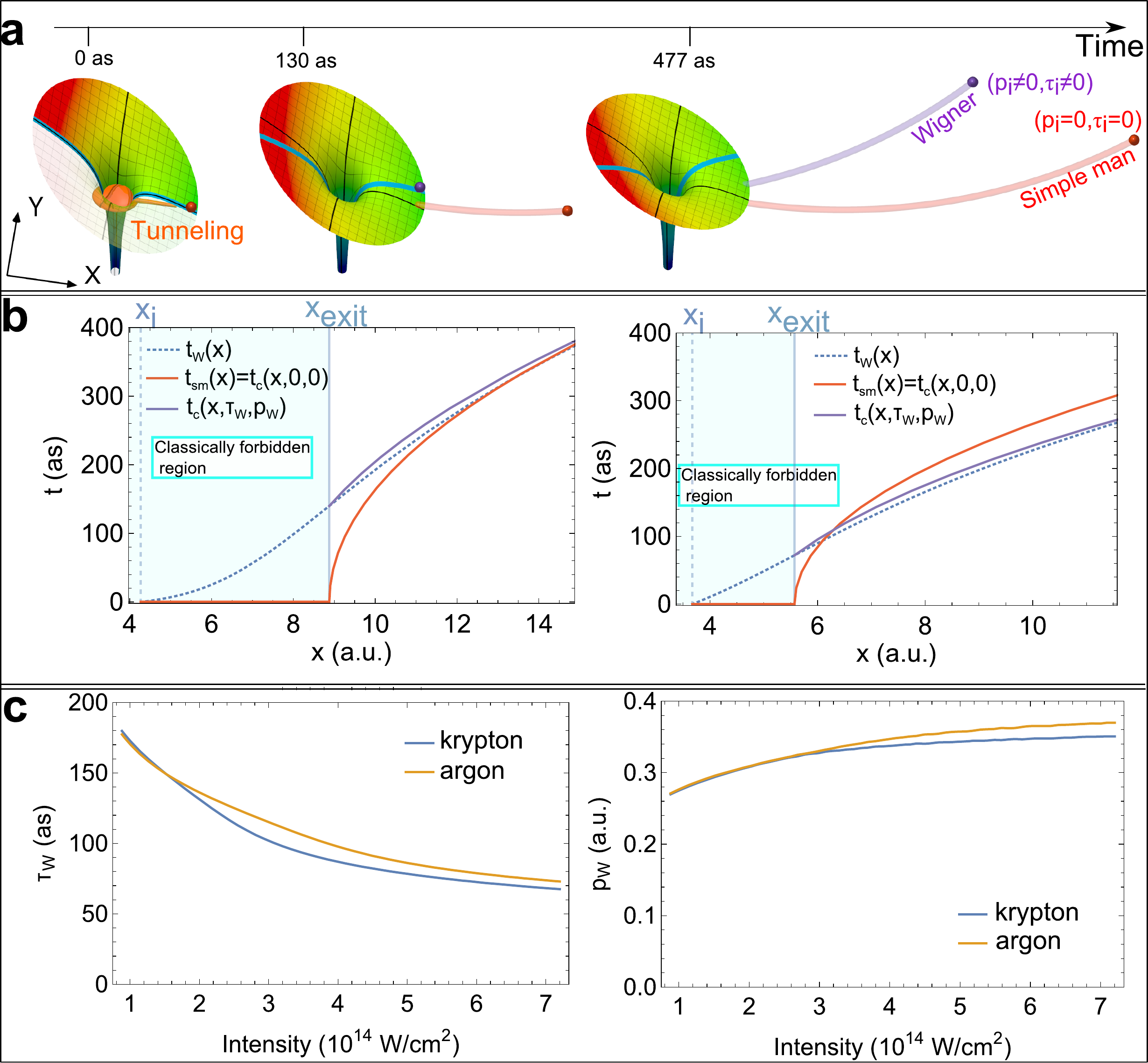}
  \caption{$\vert$ \textbf{Semi-classical picture and initial conditions provided by the Wigner formalism.} \textbf{a}, Schematic representation of the ionization picture: the Coulomb potential is bent by the laser electric field forming a saddle in the potential (blue line). The bound electron wave packet is ionized through tunneling and appears as a classical particle at the tunnel exit. Depending on the initial condition of the particle, different trajectories arise and appear with different momentum on the detector:  Simple-man model (red trajectory) and the case where the initial conditions are given by the Wigner formalism (purple trajectory). \textbf{b}, Electron trajectories in different models at low (left panel) and high (right panel) laser intensities: (dashed, blue) Wigner trajectory via Eq.~(\ref{Wigner_1}), (red, solid) the simple-man trajectory, (purple, solid) classical trajectory with the initial conditions given by the Wigner formalism. The trajectories are calculated for a krypton atom. The laser intensities are $I=1.7$ (left) and $I=6.1 \times 10^{14}~\, \mbox{W}/\mbox{cm}^2$ (right). The trajectories are shown in one-dimensional (along the tunneling direction) Cartesian coordinate. \textbf{c}, Initial conditions arising from the Wigner formalism for the two investigated atoms (argon and krypton): initial time (left panel) and initial longitudinal momentum (right panel). Notably, at the tunnel exit, the differences between both targets, argon and krypton, are as small as 10 attoseconds for the time delay and less than 0.1 a.u. for the longitudinal momentum.}
 \label{ionization_picture_1}
\end{figure}

To specifically probe the tunneling step one takes advantage of the fact that the electron, once ionized, is driven by the laser electric field. Using close-to-circularly polarized laser pulses (attoclock configuration), and therefore a rotating electric field, the instant of time when the electron appears in the continuum is effectively mapped onto a characteristic emission direction that can be measured after the pulse \cite{Eckle2008,Eckle2008a,Pfeiffer2012n,Landsman2014o} (Fig.~\ref{ionization_picture_1}a). In order to correctly disentangle the tunneling step information (time, momentum and position right after tunneling) in the final photoelectron momentum distribution, and to achieve a meaningful quantitative interpretation of attoclock results, the laser pulse parameters must be known and the Coulomb-interaction with the remaining ion along the electron excursion in the continuum must be taken into account. At present, we are witnessing an ongoing debate about the role of the initial electron momentum \cite{Barth2011,Pfeiffer2012,Li2013,Klaiber2015a,Li2016} as well as the question whether and how the electron motion can be traced back to the tunneling step \cite{Pfeiffer2012n,Landsman2014o,Ni2016}.

In our approach the tunneling dynamics is described  quantum mechanically for a quasistatic barrier via solution of the  Schr\"{o}dinger equation, and fully accounts for the Coulomb interaction  during the under-the-barrier dynamics.  For a full description of the electron dynamics in the laser field, we connect the tunneling step to the following second step of classical excursion in the continuum using the initial conditions provided by our quantum calculation (Fig.~\ref{ionization_picture_1}a), and deduce the signatures of the tunneling step that emerge in the asymptotic photoelectron momentum distribution.

\begin{figure} [ht]
\centering
\includegraphics[width=01.0\textwidth]{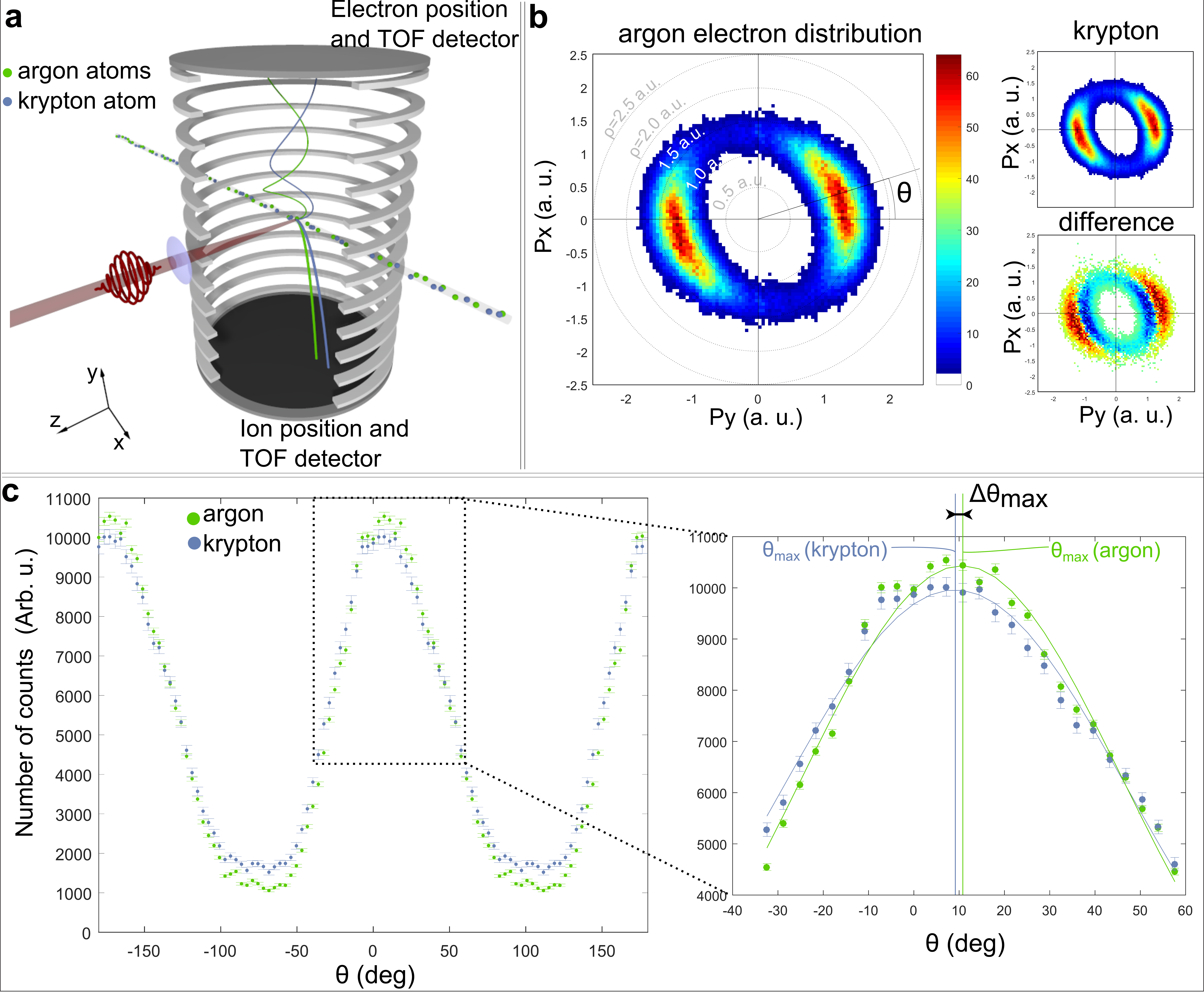}
\caption{$\vert$ \textbf{Experimental acquisition of the photoelectron emission angles.} \textbf{a}, Sketch of the experimental setup: close to circularly polarized laser pulses with a $1300\,$nm wavelength are focused onto a gas mixture of argon and krypton inside a reaction microscope. \textbf{b}, Experimental momentum distribution obtained when selecting the argon ionization events (left panel) or krypton (top right panel). The difference between the normalized distributions is indicated (bottom right). \textbf{c}, Representation of the number of counts as a function of the $\theta$ angle for both distributions from which we extract $\theta_{max}$ the angle at which the number of counts is maximum.}
\label{Experimental_strategy}
\end{figure}

The theoretical predictions are compared with the experiment, where a high sensitivity on the tunneling step is achieved by measuring strong-field tunneling ionization of atomic species with slightly different ionization potentials, but under otherwise absolutely identical laser and experimental conditions. To achieve this, we simultaneously collect within the same experiment photoelectron momentum distributions with high resolution for ionization of a gas mixture containing argon and krypton. Elliptically polarized, 60 fs laser pulses (1300 nm) at various intensities in the low $10^{14}$ W/cm$^2$ range are focused into the gas beam at the centre of a reaction microscope \cite{Ullrich2003} that allows for the coincident detection of electron-ion pairs created by ionization of Ar or Kr atoms. The electron spectra of both species are separated unambiguously by means of ion-tagging allowing comparison with each other (Fig.~\ref{Experimental_strategy}a).

Rather than the absolute value of the most probable photoelectron emission angle we analyse the angle-difference between both targets in comparison with theory. This quantity is found to be most sensitive to the initial conditions of the electron trajectory right after tunneling, and, at the same time, almost insensitive to systematic errors or any other experimental deficiencies. 

\FloatBarrier

For the first step we consider the propagation of the electron's wave function, originating from the initial bound state $\Psi(\vec{r}_i,0)$, by the space-time propagator $K(\vec{r},\vec{r}_i,t)$ (atomic units $m_e=\hbar=e=1$ are used throughout): 
\be
\Psi(\vec{r},t) =  \int d \vec{r}_i \, K(\vec{r},\vec{r}_i,t)\, \Psi(\vec{r}_i,0) \, .
\ee
From the infinite number of contributing trajectories from $\vec{r}_i$ to $\vec{r}$, we identify the most dominant one. We use quasi-static description of the laser field which is justified due to the large value of the laser period with respect to the characteristic time in the tunneling regime, described  by the smallness of the Keldysh parameter \cite{Keldysh1965}. In this case the electron passes the potential barrier with an almost constant energy. Accordingly, generalizing the well-known Wigner approach \cite{Wigner1955}, the most dominant trajectory along the tunneling channel $t_W (x)$ is determined by the phase of the fixed-energy propagator $G(x,x_i, \epsilon)$, which is the Fourier-image of $K(x,x_i,t)$:
\be
\label{Wigner_1}
t_W (x) = \left. \frac{\del \arg\left[ G(x,x_i, \epsilon) \right]}{\del \epsilon} \right|_{\epsilon= - I_p}\, ,
\ee
where $I_p$ is the atomic ionization potential including the Stark-shift  in the laser field, $x \equiv \hat{\vec{n}} \cdot \vec{r}$ is the one-dimensional tunneling coordinate with $\hat{\vec{n}}$ being the unit vector along the tunneling channel, $x_i$ is the starting point of the Wigner trajectory at time $t_0=0$, corresponding to the peak of the laser electric field, and $\arg\left[ G(x,x_i,\epsilon)\right]=S(x, \epsilon)-S(x_i, \epsilon)$ with $S(x, \epsilon)$ being the phase of the wave function of the stationary Schr\"{o}dinger equation, which is solved numerically (see details in Supplemental Material). 

Two exemplary Wigner trajectories  are shown in Fig.~\ref{ionization_picture_1}b. They both manifest that for the most probable trajectory the electron appears at the tunnel exit $x_{exit}$ not instantaneously but with a time delay   $\tau_W \equiv t_W (x_{exit}) > 0$. Moreover, we predict a non-vanishing longitudinal momentum along the laser field direction with which the electron appears in the continuum:   $p_W = \left( d \, t_W (x) / d x  \right)^{-1}|_{x = x_{exit}}$. The value of these parameters are decisive for the interpretation of the attoclock measurement and are heavily discussed \cite{Eckle2008,Pfeiffer2012,Li2013,Landsman2014o,Yakaboylu2014,Ni2016,Torlina2015}. In our approach, both, the time delay and the initial momentum are inherently provided by the Wigner trajectory $t_W(x)$. With increasing laser intensity, resulting in an effective shortening of the barrier, the tunneling time delay decreases and the longitudinal momentum becomes larger (Fig.~\ref{ionization_picture_1}c).

To connect the tunneling step with the subsequent electron motion in the classically allowed region during the second step, we define a classical trajectory starting at the tunnel exit $x_{exit}$ at time $t_{exit} = \tau_W$ with an initial momentum $p_{exit} = p_W$ that mimics the Wigner trajectory (Fig.~\ref{ionization_picture_1}c). As the laser field rotates during the tunneling time delay,  the electron emission direction as well as the position  of the tunnel exit are adapted to the field direction after the time delay $\tau_W$.
The final electron momentum after the laser pulse is calculated via classical propagation
\be
\label{clas_prop}
\vec{p}_f = \vec{p}_{exit} - \int_{t_{exit}}^\infty dt\, \left[ \vec{E}(t) + \vec{\nabla} U(\vec{r}(t)) \right] \, ,
\ee
by solving Newton's equations of motion $\ddot{\vec{r}}(t) = - \vec{\nabla} U(\vec{r}(t))-\vec{E(t)}$, with $\vec{E}(t)$ being the laser electric field and $U(\vec{r})$ the potential of the atomic core taking into account its polarization  by the laser \cite{Dimitrovski2010}.

The influence of the initial conditions on the asymptotic electron momentum is illustrated in Fig.~\ref{ionization_picture_1}b by comparing the Wigner trajectory $t_W (x)$ with the trajectory of the commonly used classical simple-man (SM) model $t_{sm} (x)$ where the electron appears instantaneously at the tunnel exit with an initial momentum which is equal to zero. Neglecting the atomic Coulomb potential in the SM model leads to the final momentum $\vec{p}_f = -  \int_{t_{0}}^\infty dt\,  \vec{E}(t)$. Accordingly, the additional time delay $\tau_{W}$ manifests as a rotation of the asymptotic momentum distribution by $\delta \theta_{\tau} = \omega \tau_{W}$, and the non-zero initial momentum to a counter-rotation by $\delta\theta_{p}\approx-p_w/p_E$ with $p_E= E_0/\omega$, with $\omega$ and $E_0$ being the laser-field frequency and amplitude, respectively. At low intensities both effects compensate each other almost completely and the corresponding trajectories merge (Fig.~\ref{ionization_picture_1}b left panel). This explains, to a large extent, the success of the classical SM model in predicting the correct electron momentum \cite{Yakaboylu2014}. However, with increasing intensity the Wigner time delay decreases faster than the initial momentum increases leading to an additional net rotation compared to the SM prediction $\delta \theta =\delta \theta_{\tau}-\delta\theta_{p}$ (Fig.~\ref{ionization_picture_1}b right panel).

To demonstrate the relevance of the initial parameters ($\tau_W$,$p_W$) and their influence on the final momentum distribution, we utilise their strong dependence on the width of the tunneling barrier.

We compare the theoretical predictions with experimental electron spectra for two atomic species with slightly different ionization potentials $I_p$. The data for both targets were collected within the same experiment for absolutely identical laser and experimental conditions (Fig.~\ref{Experimental_strategy}a). By choosing a gas-mixture of argon ($I_p ^{Ar}$=15.76 eV) and krypton ($I_p ^{Kr}$=13.99 eV), two species with similar physical and chemical properties, we can experimentally probe the barrier width and effectively reduce the  influences  of other effects such as differences in Stark-shifts, atom polarizabilities or dependences on the initial-state orbital momentum \cite{Barth2011} (see supplementary material). 

Laser pulses with a wavelength of $1300\,$nm and an ellipticity of $0.85\pm0.05$ have been emplyed. By variation of the laser intensity a detailed investigation over a large range of effective tunneling barrier widths becomes possible.

\FloatBarrier

Measured electron momentum distributions for argon and krypton are shown in Fig.~\ref{Experimental_strategy}b at a laser intensity of about $2 \times 10^{14} W/cm^2$. As expected for atoms with similar ionization potentials, the spectra for both targets appear almost identical. However, clear differences manifest after the subtraction of the normalized distributions (Fig \ref{Experimental_strategy}b). For comparison with theory, as mentioned above, the value of interest is the most probable electron trajectory, in particular the final momentum and the angle of rotation with respect to the axes of the polarization ellipse \cite{Eckle2008a}. The latter we associate with the angle $\theta_{max}$ of the distribution where the number of counts is maximum. The angle $\theta_{max}$ is the observable that is most sensitive to both time delay and initial momentum. It is obtained from a Gaussian fit to the experimental angular distribution in range from $-50^{\circ}$ to $+50^{\circ}$ (Fig~\ref{Experimental_strategy}c).

\begin{figure} [ht]
\centering
\includegraphics[width=0.8\textwidth]{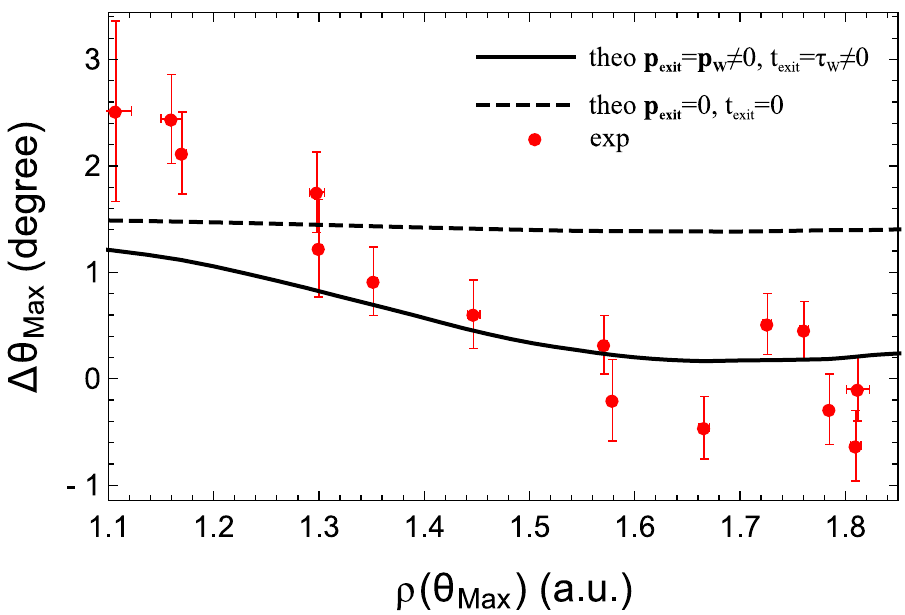}
\caption{$\vert$ \textbf{Difference between the most probable photoelectron emission angle for argon and krypton.} Experiment and theories (with and without initial momentum and tunneling delay time). Note that the indicated tunneling theory is not fully applicable to the first three data points where the intensity is in the intermediate tunneling-multiphoton regime.  
}
\label{Angle_distribution}
\end{figure}

Instead of analysing the absolute emission angles for both targets, which would require a very precise determination of the orientation of the polarization ellipse, we concentrate on the orientation independent angle difference $\Delta \theta_{max}=\theta_{max}^{Ar}-\theta_{max}^{Kr}$ and follow its behaviour as a function of intensity. Again, to circumvent the notorious difficulties in determining the laser intensity with high accuracy, we use instead of not well determined experimental intensities the average electron momentum at $\theta_{max}$ for ionization of argon: $\rho(\theta_{max})=\rho^{Ar}(\theta_{max})=\sqrt{p_x^{Ar}(\theta_{max})^2+p_y^{Ar}(\theta_{max})^2}$. For circularly and elliptically polarized pulses this average momentum $\rho(\theta_{max})$ is an unambiguous measure of the actual intensity, in which the correct mapping from momentum to intensity must be provided by theory \cite{Alnaser2004,Boge2013}. Finally, we plot $\Delta \theta_{max}$ as a function of $\rho(\theta_{max})$ and compare with our theoretical model (Fig.~\ref{Angle_distribution}).

For a conclusive comparison two additional contributions, that are unavoidable in any experiment, must be considered: the focal volume averaging and the depletion effects. The theoretical curves for two sets of initial parameters are shown in Figure~\ref{Angle_distribution}, the solid line is for the initial conditions given by the Wigner formalism, and the dashed curve for the standard conditions (simple-man model) of zero time delay and zero initial momentum.  The first observation is that - as expected- the initial conditions influence the final angle difference. Secondly, our results demonstrate that the simple-man model fails in reproducing subtle features in strong field ionization, which are very well reproduced with a more refined theory based on the Wigner formalism. The agreement with experiment remains even when varying in our model all possible parameters such as ellipticity, pulse duration, and pulse shape. There is no convincing way to explain the current experimental result without the initial longitudinal momentum and the time delay (see Supplemental Material).

\FloatBarrier

In conclusion, we have addressed the time-resolved dynamics of the tunneling ionization of an atom.  We identified theoretically the tunneling delay time and the initial longitudinal momentum at the tunnel exit for the most probable trajectory of the tunneling electron and tested it experimentally. The experimental results can be satisfactorily reproduced only when these two parameters are included in the theoretical model resulting in a non-vanishing tunneling time.

\appendix

\section{Experimental Details}

\subsection{Basic principle} \label{Basic_principle}

In order to extract information about the tunneling process we study electron momentum distributions for ionization of two atomic systems with similar ionization potential: argon and krypton. Using the coincidence capabilities of the Reaction Microscope we investigate the two atoms within the same experiment, at the same time to avoid the intrinsic experimental inaccuracies of independent measurements (changes in the laser properties and systematic imprecisions).

\subsection{Laser setup}  \label{Laser_setup}

Linearly polarized 1300nm radiation was generated by an optical parametric amplification (OPA) system (TOPAS-C, Light Conversion) pumped by $25\,fs$, $790\,nm$ pulses from a Ti:Sapphire based laser system. The close-to-circular polarization of the near-IR pulses was adjusted using broadband half and quarter wave plates. The ellipticity was determined to be $0.85\pm0.05$ (the ratio of the minor to the major axis). The pulse duration was determined to be $60\pm15\,fs$ at FWHM. These two parameters were measured using optical methods, independently from model dependent fitting of experimental photo-ionization momentum distributions. The laser intensity at the target was adjusted using a motorized iris. To ensure that no laser parameters are drifting during the measurements, the polarization and intensity were monitored in-situ by inspection of the ionization count rate and the momentum distribution of the recorded electrons. As indicated in the main text, the absolute calibration of the experimental intensity was performed by comparing the momentum distribution to the simulations including effects intrinsic to the experiment: depletion of target atoms and focal volume averaging\cite{Alnaser2004_A}.

\subsection{Spectrometer}    \label{Spectrometer}

The laser pulses are focused onto a supersonic, internally cold, gas jet inside a Reaction Microscope (ReMi) (see Fig~[2A] in the main text). The reaction fragments (electrons and ions) are guided by weak homogeneous electric and magnetic fields onto time and position sensitive detectors. From their time of flight and their impact position, the initial momentum vectors of the fragments are determined \cite{Ullrich2003_A,Jesus2004}. For the present analysis the electron spectra are utilized. For a given field setting of the spectrometer, the electron momentum transverse to the spectrometer axis is not resolved at longitudinal momenta that correspond to electrons with a time of flight equal to multiples of their cyclotron period. In order to account for this, spectra at different field settings were recorded and combined. 

We used a gas mixture of argon and krypton. Using the coincidence with the ion for assigning electrons coming from argon or krypton atoms we were able to measure the 3D electron-momentum distribution for each target maintaining identical experimental conditions meaning with the same laser pulse parameters and the same spectrometer settings. For the given intensities, the count rate was less than one ionization event per pulse leading to false coincidences of less than $4\%$.

\subsection{Data analysis}  \label{Data_analysis}

For each momentum distribution, a peak search is performed to determine the exact angle at which the maximum number of counts is observed (see also main text). To do so we fit a Gaussian function to the peak in the angular distribution from -50 to +50 degree with the amplitude, the peak position and the width being free parameters. The fit is performed on the data weighted by their errors. The error bars shown on the angle difference (see Fig [3] in the main text) represent the one sigma statistical error obtained through a standard error propagation. A second method is used to verify this obtained angle difference between the two species. A cross-correlation between the angular distribution curves of the two species leads to the same value within the error bars. 

The observable $\rho(\theta_{max})$ used in the following and in the main text correspond to the average radial electron momentum at the peak of the angular distribution for the case of argon $\sqrt{p_x^{Ar}(\theta_{max})^2+p_y^{Ar}(\theta_{max})^2}$. It is used as a measure for the laser intensity. Since the conversion between intensity and the radius of the 2D momentum distribution is disputable \cite{Ivanov2014,Boge2013_A}, we refer to this experimental observable as it is extracted from the data.

\section{Theoretical Details}

In this theoretical section we start with the simplified so-called simple-man picture \cite{Simpleman2,Kuchiev1987,Smith1988,Gallagher1988,Schafer1993,Corkum1993}, which is based on quasi-classical trajectories, and then step by step we explore necessary modifications and implement them.

\subsection{Simplified picture}

We investigate tunnel-ionization in a regime where the Keldysh parameter is small $\gamma \quad = \omega \sqrt{2I_p}/E_0  < 1$, with $\omega$, $E_0$, and $I_p$ being the laser frequency, the electric field amplitude, and the ionization potential, respectively. Additionally, the condition $E_0/E_a \lesssim 1/9$, with the atomic field strength $E_a =(2I_p)^{3/2}$, is fulfilled and there is no over-the barrier-ionization \cite{Landau_3,pfeiffer_2012a}.

We aim to calculate the most probable asymptotic momentum of an electron liberated from an atom by strong-field tunnel-ionization. The quantum theory describes the tunneling step of the electron dynamics (see below) and provides the coordinate of the electron at the tunnel exit $\vec{r}(t_i) = \vec{r}_{exit}$, and its momentum $\dot{\vec{r}}(t_i) = \vec{p}_i $ (the dot denotes the derivative with respect to time $t$). In the continuum the electron is propagated classically  by Newton's equations
\begin{eqnarray}
\ddot{\vec{r}}(t) = -\vec{E}(t)- \bm{\nabla} U(\vec{r}(t)),
\end{eqnarray}
yielding the most probable final momentum of the photoelectron
\be
\vec{p}_f = \vec{p}_i - \int_{t_i}^\infty dt\, \left[ \vec{E}(t) + \bm{\nabla} U(\vec{r}(t)) \right] \, .
\ee

In the simple-man picture the initial momentum at the tunnel exit $\vec{p}_i$ is assumed to be zero, and the most probable instant of ionization $t_i$, corresponding to the maximum of ionization probability, is set to the time when the electric field reached its maximum value $t_{max}$. The tunneling step itself is assumed to happen instantaneously. Furthermore, the potential of the atomic core $U(\vec{r})$ is neglected, such that the liberated electron propagates freely in the electric field $\vec{E}(t)$. Then, the final momentum is determined by the vector potential $\vec{A}(t)$ at the electric field's maximum
\be
\vec{p}_f =  - \int_{t_{max}}^\infty dt\,  \vec{E}(t) =  -\vec{A}(t_{max}) \, .
\ee
The electric field of the laser propagating along the $z$-axis, we define in the non-relativistic regime
\be
\vec{E}(t) = -\mathcal{G}(t) \left[ \cos(\omega t) \hat{\vec{x}} + \zeta \sin(\omega t) \hat{\vec{y}} \right] \, ,
\ee
where $0 \le \zeta \le 1 $ is the ellipticity parameter, $\mathcal{G}(t) = E_0 f(t)$ with $E_0 = \sqrt{I/(1+\zeta^2)} $, and  $f(t)$ being the field envelope such that $f(-\infty) = f(\infty) =0 $. The intensity $I$ is given in terms of the atomic units $( 1 \, \mbox{a.u.} = 3.509 \times10^{16} \, \mbox{W}/\mbox{cm}^2)$. Here, we neglect the spatial dependency of the laser field, i.e., we apply the electric dipole approximation [see Ref.~\cite{Klaiber_2007,Klaiber_2013c} for a relativistic description of the strong field ionization  process].

In order to relate and compare experiment with theory, it is most common to use as observables the most probable electron emission angle $\theta_{max}$ and the radial momentum $\rho$ at this angle. They are given by the following formulas:
\be
\theta_{max} = \arctan\left( \frac{p_{f,x}}{p_{f,y}}\right) \, , \quad \rho(\theta_{max}) = \sqrt{p_{f,x}^2 + p_{f,y}^2} \, ,
\ee
respectively. Note that for a non-circularly polarized laser field ($\zeta < 1$) the tunneling direction for the electron contributing to the peak of the momentum distribution is defined along the $x$-axis. In the simple-man model, this yields $p_{f,x} = 0$, $\theta_{max}=0$,  and  $\rho(\theta_{max}) = \zeta E_0 / \omega $. Thereby, there is no dependence on the ionization potential and the difference in angle $\theta_{max}$ and in radius $\rho$ between different species at the same intensity would vanish.

The modifications implemented to the simple-man picture in order to reproduce the present experimental data are explained in the following sections. As a first point, we carry out averaging over the laser focus and find the focal volume averaged most probable momentum.
Next, we include the depletion and saturations effect of the ionization. Further, we include the effect of the atomic potential on the tunneling dynamics, and finally, based on quantum calculations we reveal the tunneling time and the corresponding electron momentum right after tunneling.

\subsection{Focal volume averaging} \label{focal_av}

In the experiment, ionization takes place in the entire focal volume and, accordingly, atoms at different positions in the target interact with different laser intensities.

\begin{figure}
  \centering
  \includegraphics[width=0.49\linewidth]{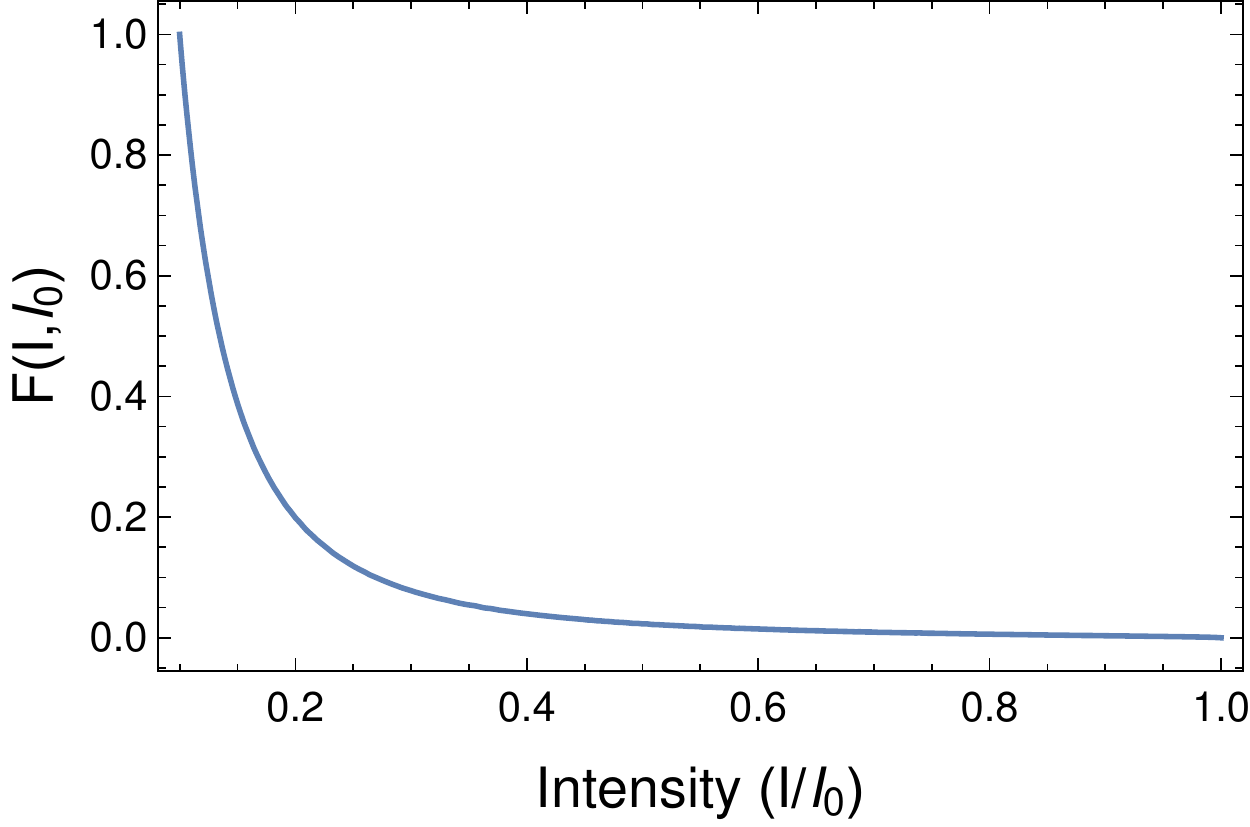}
  \caption{The distribution function of intensities $I$ that atoms experience in the laser focal volume of a Gaussian beam; $I_0$ is the laser peak intensity.}
  \label{focal_dist}
\end{figure}

The distribution function $F(I,I_0)$ of intensities $I$ that atoms experience in the laser focal volume of a Gaussian beam with peak intensity $I_0$ is presented in Fig.~\ref{focal_dist}. 
The atoms are exposed to a broad distribution of intensities up to the maximum intensity $I_0$.
Taking into account the intensity dependent ionization yield $Y(I)$, the final focus averaged most probable momentum $\bar{\vec{p}}_f (I_0)$ is then calculated by
\be
\label{focal_mom}
\bar{\vec{p}}_f (I_0) = \frac{\int_0^{I_0} d I\, F (I,I_0) Y(I) \, \vec{p}_f (I)}{\int_0^{I_0} d I\, F(I,I_0) Y(I)} \, ,
\ee
where $\vec{p}_f (I) $ is the momentum obtained for one discrete intensity $I$. The combined term $F (I,I_0) Y(I)$ can be regarded as the effective probability density for ionization which is a function of intensity $I$ at a given peak intensity $I_0$.

\subsection{Ionization yield and saturation}

In an experiment the number of atoms in the interaction region is finite, and in the laser pulse, due to ionization, the number of atoms is depleted. Therefore, some of them do not survive up to the peak of the laser pulse. We account for this depletion and saturation of ionization assuming that the most probable ionization of an atom takes place not at the  maximum of the pulse envelope but at earlier times for high intensities.

The ionization yield can be defined in terms of the instantaneous ionization rate $\Gamma(t)$ in the following way~\cite{Becker_2001,tong2005empirical}
\be
\label{yield_0}
\tilde{Y}(t) = 1 - \exp\left[- \int_{-\infty}^t d t' \, \Gamma(t') \right] \, .
\ee
In order to calculate the instantaneous ionization rate we replace $E_0$ with $\sqrt{E_x^2 (t) + E_y^2 (t)}$, and the Keldysh parameter $\gamma$ with $\gamma(t)\equiv  \omega \sqrt{2 I_p}/\mathcal{G}(t)$ in the well-known ADK formula with the leading $\gamma$ correction~\cite{PPT2,PPT3}
\bal
\label{rate}
& \Gamma (t) =  C_l^2 \frac{3 \exp\left[ - \frac{9}{I_p}\frac{\mathcal{G}(t)}{(2I_p)^{3/2}} \right] }{2 (2 I_p)^{1/\sqrt{2I_p}-1/2}} \left(\frac{2 (2I_p)^{3/2}}{\sqrt{E_x^2 (t) + E_y^2 (t)}}\right)^{\frac{2}{\sqrt{2 I_p}}-1} \\
\nonumber & \times \exp\left[ - \frac{2 (2I_p)^{3/2}}{3 \sqrt{E_x^2 (t) + E_y^2 (t)}}  \left(1 - \frac{1}{10} \left(1 - \frac{\zeta^2}{3} \right)\gamma(t)^2 \right) \right] \, .
\eal
The exponent in the first line is an empirical correction~\cite{tong2005empirical} with ${C_l}_{Ar} = 2.44$ and ${C_l}_{Kr} = 2.49$~\cite{tong2002theory}.

\begin{figure}
  \centering
  \includegraphics[width=0.49\linewidth]{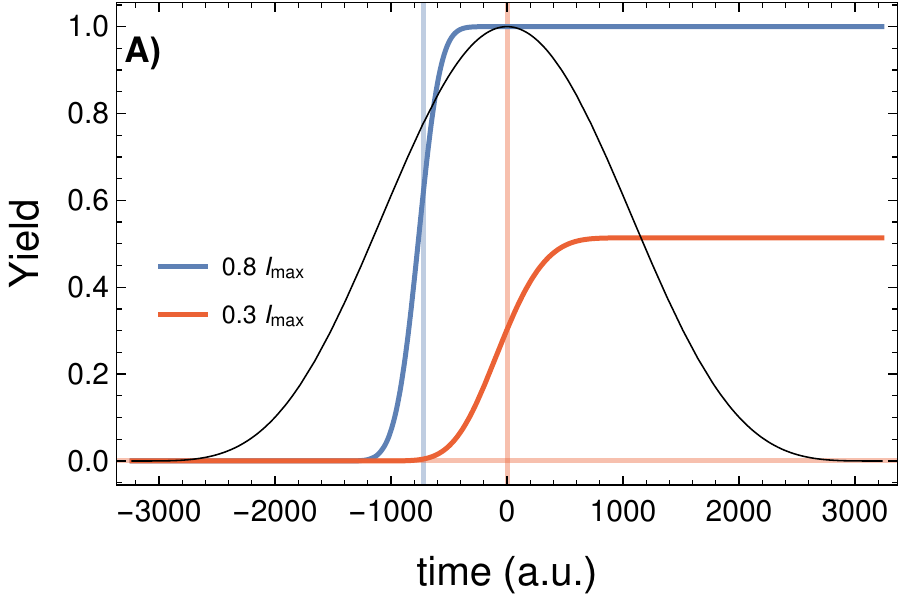}
    \includegraphics[width=0.49\linewidth]{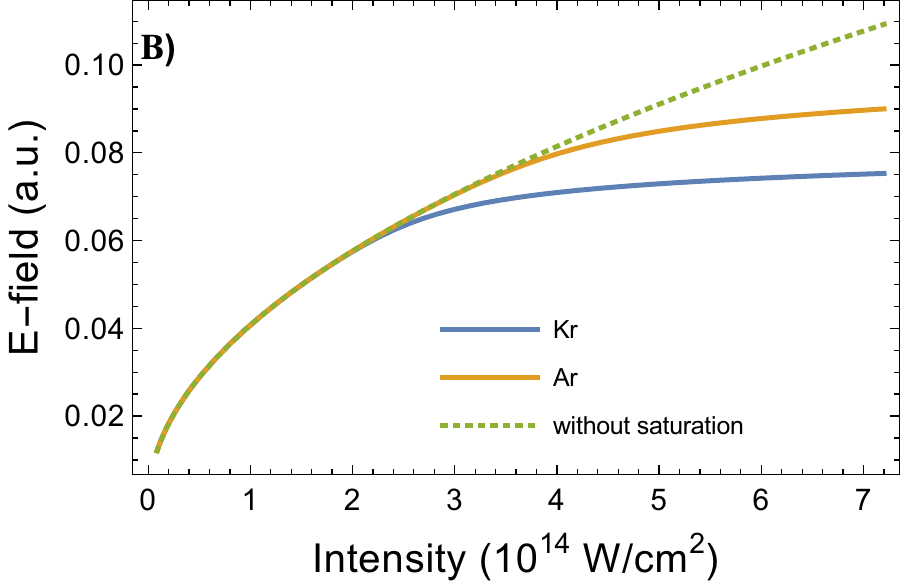}
  \caption{Figure A) shows the yield for Kr as a function of time in a laser pulse (black thin solid line). For high intensities the yield saturates before the peak of the electric field. The vertical lines are the corresponding saturation times. In figure B) the electric field seen by Kr (blue solid curve), and Ar (yellow solid curve) are shown, the dotted green curve shows the field without the saturation effect. The applied parameter are $\omega = 0.0350$, $\zeta = 0.85$, $I_{max} = 7.2 \times 10^{14} \, \mbox{W}/\mbox{cm}^2$ and the envelope is $\cos^4 (t \pi / \tau )$ with $\tau = 36 \times 2 \pi / \omega = 55 \, \mbox{fs}$.}
  \label{yield_sat}
\end{figure}

The yield at time $t$ via Eq.~\ref{yield_0} is presented in Fig.~\ref{yield_sat} A, which shows that the yield saturates before the electric field reaches its peak value for high intensities. In other words, all the atoms are already ionized before they have a chance to interact with the peak intensity. 
Most atoms are ionized at the moment $t_s$ when the yield derivative is maximal, i.e., at $\ddot{\tilde{Y}}(t_s) = 0$. Therefore, we conclude that the most probable electron trajectory starts its classical motion in the continuum not at the field maximum, but at an earlier time $t_s \le t_{max}=0$.
As a result, the electric field observed by the liberated electron saturates with increasing intensity, see Fig.~\ref{yield_sat} B. The yield needed for the focal averaging in Eq.~(\ref{focal_mom}) is then given by $Y(I) = \tilde{Y}(t_s)$. 

The saturation time $t_s$ is an important parameter which has an impact on the most probable phototelectron momentum. At first sight the derivation $t_s$ is a critical point of the analysis, because it depends on  many factors (the intensity, rate formula, the ellipticity, pulse duration) which are predetermined only approximately in the experiment. However, we will show below that the main observables, the difference of the angle defining the most probable photoelectron momentum for argon and krypton, is robust against variations of the mentioned parameters.

\subsection{Binding potential, ionization energy, and tunnel exit}

Both the tunneling step and the electron dynamics in the continuum are governed not only by the electric field, but also by the potential of the atomic core. The binding potential of a multi-electron atom can be modelled with a Hartree-Fock-Slater type potential, which is split into two parts; the usual Coulomb potential, and a screening potential, that reflects the multi-electron effect,
\be
U(\vec{r}) = - \frac{1}{r} - \frac{\Phi(r)}{r} \, ,
\ee 
where the screening part $\Phi(r)$ vanishes for large values of $r$. The screening function is usually written in the following form \cite{Muller_99,Cloux_15}:
\be
\Phi(r) = A \exp(- B r) + (Z-1-A)\exp(-C r) \, ,
\ee
with $Z$ being the atomic number, and the coefficients $A$, $B$, and $C$ are chosen such that the bound state energies fit with the experimental ones. For argon ($Z=18$) these coefficients are $A_{Ar}= 5.4$, $B_{Ar} = 1$, and $C_{Ar} = 3.682$ \cite{Muller_99}, whereas in the case of krypton ($Z=36$) the coefficients are $A_{Kr}= 6.42$, $B_{Kr} = 0.905$, and $C_{Kr} = 4.20$ \cite{Cloux_15}.

\begin{figure}
  \centering
  \includegraphics[width=0.49\linewidth]{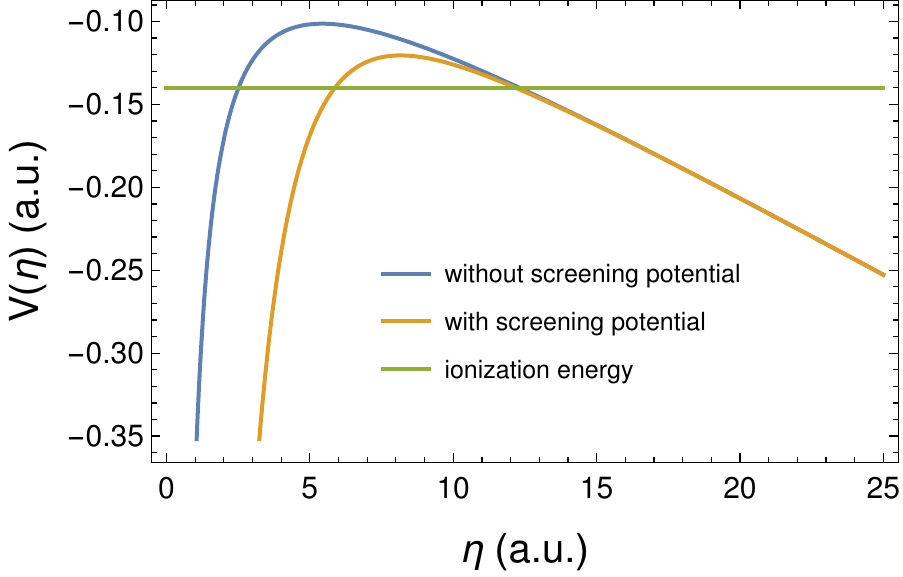}
  \caption{The potential barrier~(\ref{pot_barrier}) as a function of $\eta$. In the potential the polarization term is omitted in order to make the difference visible. The applied parameters are $t_s = 0$, $I = 0.5 I_{max}$ with the krypton parameters.}
  \label{barrier1}
\end{figure}

Furthermore, due to the interaction with the electric field the atom is polarized. This affects the ionization potential via a Stark shift
\be
I_p = I_p^0 + \frac{1}{2}(\alpha_N - \alpha_I) \mathcal{G}^2(t_s) \, ,
\ee
where $I_p^0$ (0.580 a.u. for argon and 0.515a.u. for krypton) is the field-free ionization energy, $\alpha_N$ and $\alpha_I$ are the static atomic and ionic polarizabilities, respectively \cite{pfeiffer_2012a}. The corresponding static polarizabilities are $\alpha_{N,Kr} = 16.7 $, $\alpha_{I,Kr} = 9.25 $ , $\alpha_{N,Ar} = 11.1 $, $\alpha_{I,Kr} = 7.2 $ \cite{shevelko1979static,schwerdtfeger2014table}. The polarization also affects the potential of the atomic core \cite{Dimitrovski_10} 
\be
U(\vec{r}) = - \frac{1}{r} - \frac{\Phi(r)}{r} - \alpha_I \frac{\vec{r} \cdot \vec{E}(t_s)}{r^3} \  \, .
\ee 
After all, the ionized electron propagates in an effective potential
\be
U(\vec{r},t) = - \frac{1}{r} - \frac{\Phi(r)}{r} - \alpha_I \frac{\vec{r} \cdot \vec{E}(t_s)}{r^3}  + \vec{r} \cdot \vec{E}(t) \\  \, 
\ee 
after the tunneling. 

In order to find the position of the tunnel exit, we first write down the corresponding time independent Schr\"{o}dinger equation (TISE) with the potential at time $t_s$
\be
\label{tise_0}
-I_p \psi(\vec{r}) = \left( - \frac{\bm{\nabla}^2}{2} - \frac{1}{r} - \frac{\Phi(r)}{r} + \alpha_I \frac{x\, \mathcal{G}(t_s)}{r^3}  - x\, \mathcal{G}(t_s) \right ) \psi(\vec{r}) \, .
\ee
In parabolic coordinates, $x = (\eta - \xi)/2$, $y = \sqrt{\eta \xi} \cos(\phi)$, $z = \sqrt{\eta \xi} \sin(\phi)$, Eq.~(\ref{tise_0}) can be separated~\cite{Landau_3,bisgaard2004tunneling,pfeiffer_2012a}, and in the limit $\eta \gg \xi$, we find the one-dimensional potential barrier
\be
\label{pot_barrier}
V_{B}(\eta) = - \frac{1 - \sqrt{2 I_p}/2}{2 \eta} - \frac{1}{8 \eta^2} - \frac{\mathcal{G}(t_s) \eta}{8} + \alpha_I \frac{\mathcal{G}(t_s)}{\eta^2} - \frac{\Phi(\eta/2)}{2 \eta} \, ,
\ee
The tunnel exit in parabolic coordinates is then given by $V_{B}(\eta_{exit}) = - I_p /4$, which in Cartesian coordinates reads $x_{exit} = \eta_{exit}/2$. Although the screening potential has a negligible effect on the tunnel exit, see Fig.~\ref{barrier1}, its impact is crucial for the quantum mechanical propagation, which we discuss in the next section.

\subsection{Tunneling time and initial tunneling momentum}

The Schr\"odinger equation (\ref{tise_0}) corresponds to the tunnel-ionization problem. When described within the quasi-classical (Wentzel-Kramers-Brillouin (WKB)) theory the
tunneling is an instantaneous process, and the tunnel exit is a turning point, where the electron is ionized with a vanishing longitudinal initial momentum. In order to reproduce a more accurate tunneling picture we go beyond the WKB approximation in our quantum mechanical description.

The quantum mechanical description of the propagation of the ionized electron is given by the propagation of the wave function
\be
\Psi(\vec{r},t) = \bra{\vec{r}} U(t,t_s) \ket{\Psi (t_s)} = \int d \vec{r}_i \, K(\vec{r},\vec{r}_i;t,t_s) \Psi(\vec{r}_i,t_s) \, ,
\ee
where $\Psi(\vec{r}_i,t_s)$ is the initial wave function, and the term $ K(\vec{r},\vec{r}_i;t,t_s) = \bra{\vec{r}} U(t,t_s) \ket{\vec{r}_i}$ is the corresponding space-time propagator, or Feynman Kernel, which connects space points $\vec{r}_i$ and $\vec{r}$ in a time interval $t-t_s$.

The Keldysh condition ($\gamma < 1$) guarantees that the ionization process is adiabatic. Therefore, we can assume that the energy is conserved during the process, and accordingly the space-time propagator can be written in terms of the fixed-energy propagator $G(\vec{r},\vec{r}_i,\varepsilon)$, the retarded Green's function of the TISE. After mapping the three-dimensional geometry to a one-dimensional tunneling geometry in the parabolic coordinates, the space-time propagator is written as
\be
K(\eta,\eta_i, t-t_s) = \frac{1}{2 \pi} \int_{-\infty}^\infty d \varepsilon \, e^{- i \varepsilon (t-t_s)} G(\eta,\eta_i, \varepsilon) \, ,
\ee
where the Green's function is given by
\be
\left(\frac{\varepsilon}{4} + \frac{1}{2}\frac{\del^2}{\del \eta^2} - V_B (\eta)  \right)G(\eta,\eta_i, \varepsilon) = \delta(\eta-\eta_i)\,.
\ee
If we decompose the Green's function into its phase and amplitude, by means of the stationary phase analysis we deduce that solutions of the equation
\be
(t-t_s) -  \frac{\del \arg\left[ G(\eta,\eta_i, \varepsilon) \right]}{\del \varepsilon} = 0
\ee
yield the dominant energies. In other respect, as in the case of tunnel-ionization, if the ionization energy is the dominant energy then the trajectory
\be
\label{Wigner_0}
t(\eta,\eta_i) = \left. \frac{\del \arg\left[ G(\eta,\eta_i, \varepsilon) \right]}{\del \varepsilon} \right|_{\varepsilon= -I_p} + t_s
\ee
will be the dominant trajectory for the space-time propagator.

This definition of the dominant trajectory is based on Ref.~\cite{Yakaboylu_2014_td}. We will call it Wigner trajectory as it generalizes the Wigner approach for the scattering time delay~\cite{Wigner_1955} to the tunneling problem. In the original idea of Wigner the asymptotic time delay at scattering was described by the derivative of the scattering phase shift with respect to energy. The definition of the dominant trajectory is quite natural. In fact, once the Green's function is calculated within the WKB approximation, the phase will correspond to the classical action, hence the definition of a trajectory by Eq.~(\ref{Wigner_0}) would follow the Hamilton-Jacobi theory (see for instance Ref.~\cite{Schulman}). 

For a one-dimensional problem the Green's function can be calculated via
\be
G (\eta,\eta_i, \varepsilon) = \frac{2 i}{W} \left[ \theta(\eta - \eta_i) \psi_{+} (\eta,\varepsilon) \psi_{-} (\eta_i, \varepsilon) + \theta(\eta_i-\eta) \psi_{-} (\eta,\varepsilon) \psi_{+} (\eta_i, \varepsilon) \right]
\ee
with $W =  \psi_{-} (\eta, \varepsilon) \del_\eta \psi_{+} (\eta,\varepsilon) - \psi_{+} (\eta,\varepsilon) \del_\eta \psi_{-} (\eta, \varepsilon)$ being the Wronskian \cite{byron2012mathematics,arfken2013mathematical,morse1953methods}, where $\psi_{+} (\eta,\varepsilon) $, and $\psi_{-} (\eta,\varepsilon) $ are the corresponding solutions of the TISE 
\be
\label{tise_eta}
\left( - \frac{1}{2}\frac{d^2}{d \eta^2} + V_B (\eta) \right) \psi_{\pm}(\eta, \varepsilon) = \frac{\varepsilon}{4} \psi_{\pm}(\eta, \varepsilon)
\ee
with positive and negative current, respectively. The phase of the Green's function can then be written as
\be
\label{phase_of_Greens}
\arg\left[ G (\eta, \eta_i , \varepsilon) \right] = S_{+} (\eta , \varepsilon) - S_{+} (\eta_i , \varepsilon)\, , \quad \eta \ge \eta_i \, ,
\ee
where $S_{\pm} \equiv \arg(\psi_{\pm})$, and we use the fact that $S_{-} (\eta_i , \varepsilon) = - S_{+} (\eta_i , \varepsilon)$. Thereby, the Wigner trajectory is defined by the phase of the wave function of the TISE with a positive outgoing probability current
\be
\label{Wigner}
t_W (\eta,\eta_i) =  \frac{\del S (\eta , -I_p)}{\del \varepsilon} - \frac{\del S (\eta_i , -I_p)}{\del \varepsilon} + t_s \, ,
\ee
where we omit the sub-index $+$ for the phase $S_{+}$.

\begin{figure}
  \centering
    \includegraphics[width=0.49\linewidth]{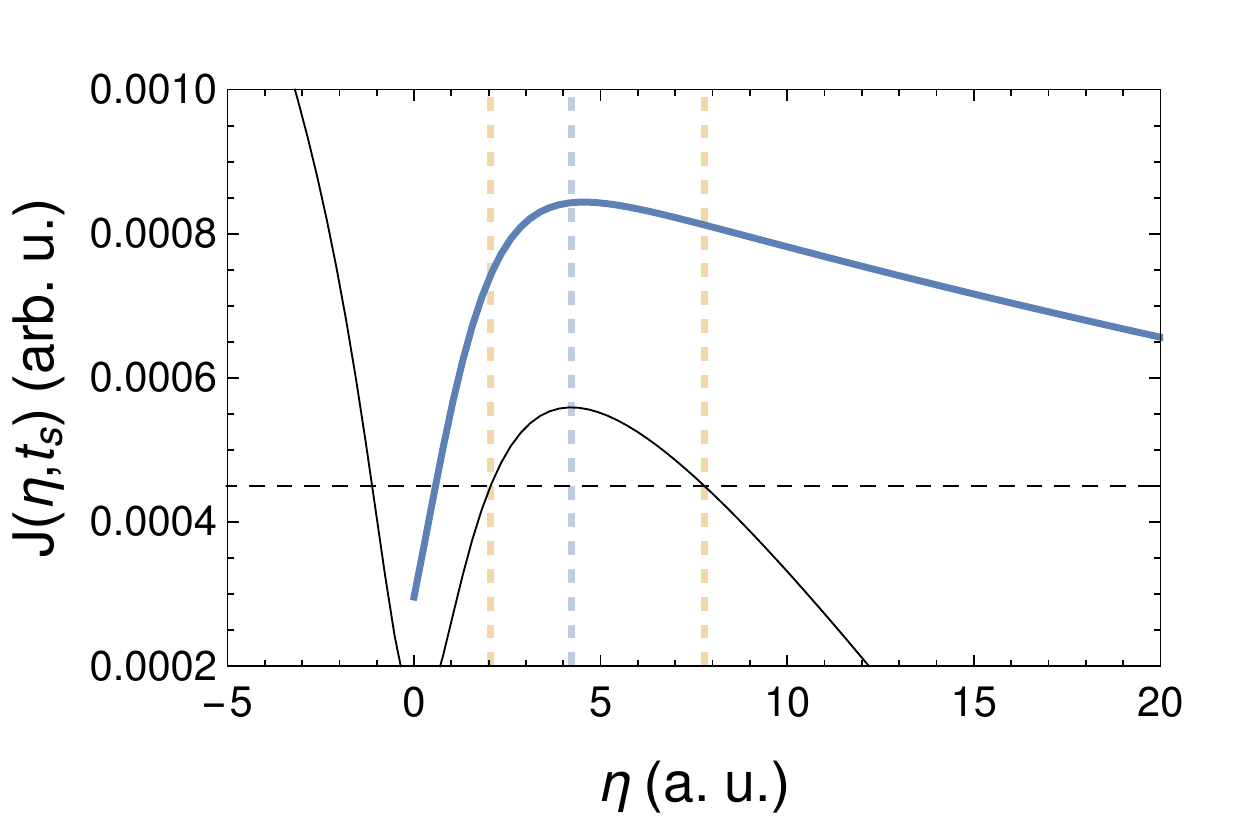}
  \caption{The probability current (blue solid line) as a function of the coordinate $\eta$ at $t=t_s$. The yellow dashed lines are the tunnelling entry and exit, respectively. The blue dashed vertical line indicates the saddle of the potential, $V_{B}'(\eta_i,t_s) = 0$. The black thin curve and the black dashed line demonstrate the tunneling barrier, and the ionization energy, respectively.}
  \label{initial_point}
\end{figure}

In order to derive the Wigner trajectory one has to specify the initial coordinate $\eta_i$. First of all, as Eq.~(\ref{Wigner}) implies $t_W (\eta_i,\eta_i) = t_s $, we conclude that the initial starting point of the Wigner trajectory $\eta_i$ is the position where the tunneling part of the electron wave packet is maximum at time $t_s$. To identify $\eta_i$ we calculate the probability current density $J(\eta,t)$ and find its maximum at $t=t_s$. In other words $\eta_i$ is given by the condition $J'(\eta_i,t_s) =0$, where the prime denotes the derivative with respect to the coordinate $\eta$. For a wave function in the form of $\psi(\eta,t) = A(\eta,t) \exp\left[ i S(\eta,t) \right]$ the probability current can be written as $J(\eta,t) = A(\eta,t)^2  S'(\eta,t)$. Using the Schr\"{o}dinger equation as well as the continuity equation we obtain
\be
S'(\eta,t) = \sqrt{2 \left(-\dot{S}(\eta,t) - V_{B}(\eta,t) + \frac{A''(\eta,t)}{2 \, A(\eta,t)} \right)}\,,
\ee
where the dot represents the derivative with respect to time and $V_{B}(\eta,t)$ is given by Eq.~(\ref{pot_barrier}) with the replacement of $\mathcal{G}(t_s) \to \mathcal{G}(t)$. We calculate the current derivative
\be
J'(\eta,t) =J(\eta,t) \left[ \frac{2A'(\eta,t)}{A(\eta,t)} + \frac{S''(\eta,t)}{S'(\eta,t)}\right]  \, .
\ee
In the quasi-static approach $\dot{S}'(\eta,t_s) \sim 0$ and in the quasi-classical case $|A''(\eta,t_s)| \ll |A(\eta,t_s)|$. Moreover, in the latter case $|A'/A|\sim |S''/S'|\sim |V_{B}'/( V_{B}+ \dot{S})| $ and one can deduce that
\be
J'(\eta_i,t_s) \propto V_{B}'(\eta_i,t_s) = 0 \, .
\ee
Namely, the probability current will be maximum at the point where $V_{B}'(\eta_i,t_s) = 0$. 
In order to justify the derived result we also solved numerically the time dependent Schr\"{o}dinger equation (TDSE) for a soft core potential with the typical parameters used in the experiment. As it is shown in Fig.~\ref{initial_point} the probability current is maximum at the saddle point of the potential $V_{B}(\eta,t_s)$.

\begin{figure}
  \centering
    \includegraphics[width=0.49\linewidth]{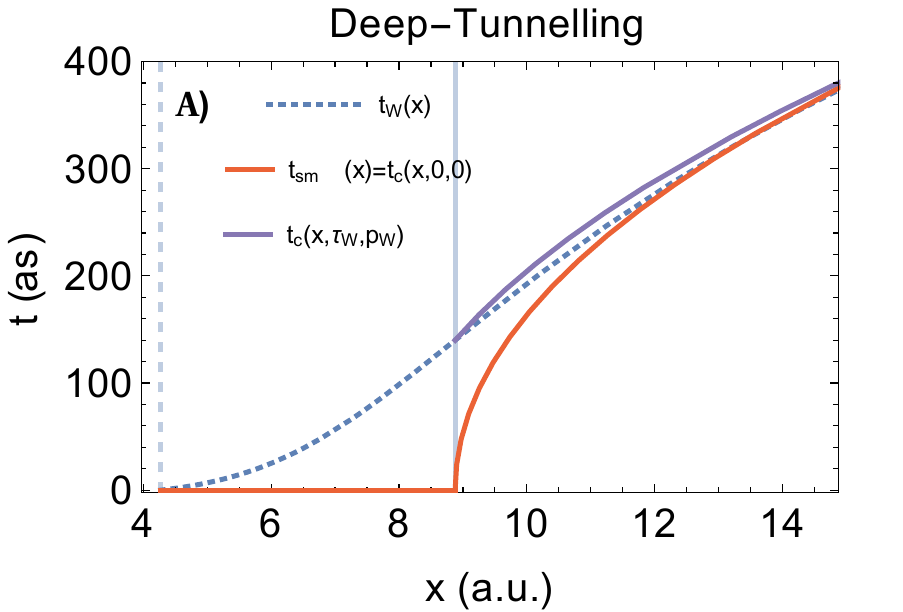}
  \includegraphics[width=0.49\linewidth]{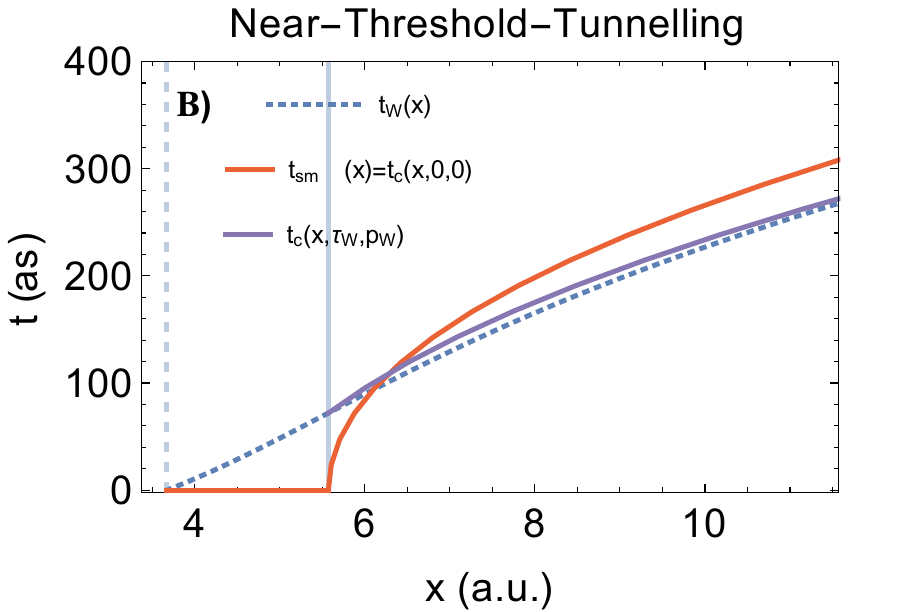}
  \caption{The Wigner (blue dotted curve), the simple-man (red solid curve), and the quasi-classical with initial condition given by the Wigner formalism (purple solid curve). The dashed vertical line corresponds to $x_i$, whereas the next solid line is the tunnel exit $x_{exit}$. The applied parameters are $I= 0.15 I_{max}$ for deep-tunneling, figure A), $I= 0.85 I_{max}$ for near-threshold-tunneling, figure B), with the krypton parameters.}
  \label{trajectories}
\end{figure}

Next, we define the Wigner trajectory along the $x$ coordinate. 
We first write the TISE Eq.~(\ref{tise_eta}) in the $x$-coordinate using the approximate coordinate transformation $\eta \approx 2 x$
\be
\left( - \frac{1}{2}\frac{d^2}{d x^2} + 4 V_B (2 x) \right) \tilde{\psi}_{\pm}(x, \varepsilon) = \varepsilon \,  \tilde{\psi}_{\pm}(x, \varepsilon) \, ,
\ee
with $4 V_B (2 x) $ being the potential barrier in Cartesian coordinates through which the electron tunnels with the energy $-I_p$. Then, we define the Wigner trajectory as
\be
\label{Wigner_x}
t_W (x) \equiv t_W (x,x_i) =  \frac{\del \tilde{S} (x, -I_p)}{\del \varepsilon} - \frac{\del \tilde{S} (x_i , -I_p)}{\del \varepsilon} + t_s \, ,
\ee
where $\tilde{S} \equiv \tilde{S}_{+} =\arg(\tilde{\psi}_{+})$ and $x_i = \eta_i /2$ as $4 V_B'(2 x_i) = 0 $.

\begin{figure}
  \centering
    \includegraphics[width=0.49\linewidth]{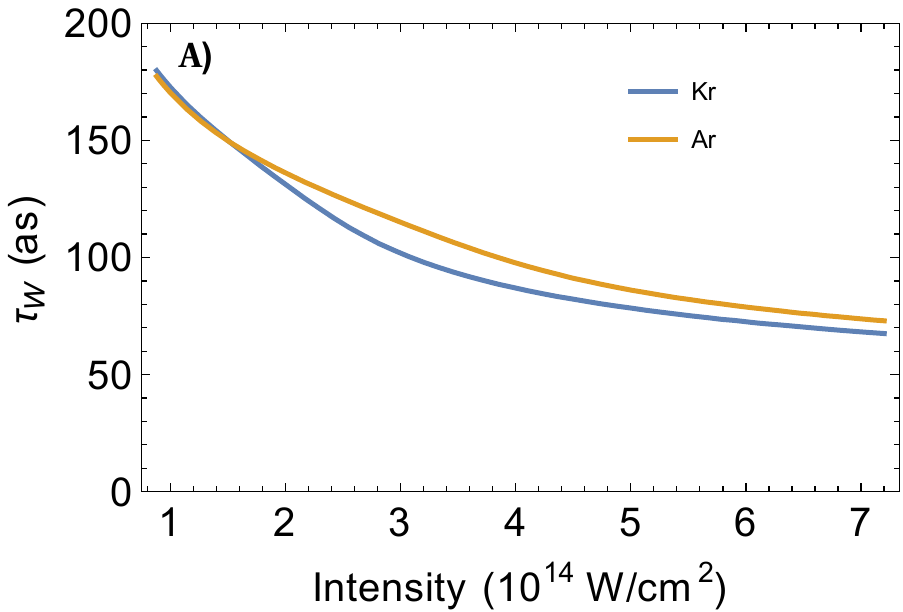}
  \includegraphics[width=0.49\linewidth]{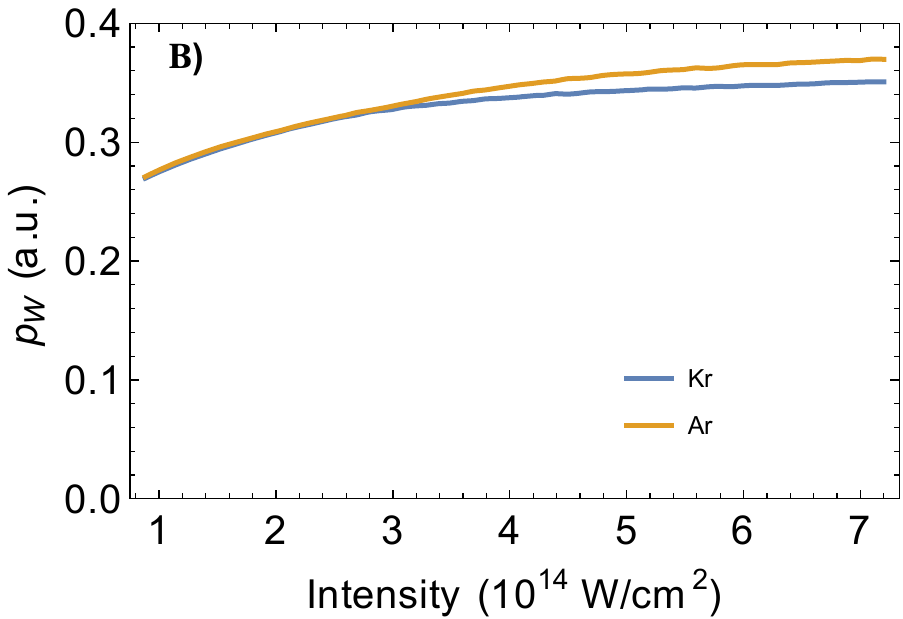}
  \caption{Figure A) shows the time spent under the barrier from $x_i$ to $x_{exit}$ as a function of intensity for Kr and Ar, and the corresponding momentum of the Wigner trajectory at the tunnel exit is shown in figure B).}
  \label{wigner_pairs}
\end{figure}

The Wigner trajectories for the two different regimes, the deep-tunneling and the near-threshold-tunneling, are shown in Fig.~\ref{trajectories}. We observe that in both regimes the electron spends a finite time under the barrier until it reaches the exit and it is set into the continuum with a positive time delay $\tau_W \equiv t_W (x_{exit}) > 0$ (see the blue dotted curves in Fig.~\ref{trajectories}). In Fig.~\ref{wigner_pairs} A) we show this time as a function of intensity for each target. Thus, the quantum mechanical treatment of the problem reveals a time delay. In addition to the time delay, the Wigner trajectory also uncovers that the electron appears in the continuum with an initial momentum $p_W = \left( d \, t_W (x) / d x |_{x = x_{exit}} \right)^{-1}$(see Fig.~\ref{wigner_pairs} B)).

Furthermore, we can compare the Wigner trajectory $t_W (x)$ with the simple-man trajectory $t_{sm} (x)$. By the latter we imply that the electron instantaneously passes the barrier and emerges at the tunnel exit with a vanishing initial momentum and obeys Newton's equations afterwards. For low intensities, which correspond to the so-called deep-tunneling regime~\cite{Yakaboylu_2013_rt}, we observe that even though the Wigner trajectory spends a non-zero finite time under the barrier, it has a non-zero initial momentum in contrast to the simple-man trajectory such that both trajectories overlap in the classical domain at distances far from the tunnel exit and hence they are indistinguishable via their asymptotic momentum distribution ~\cite{Yakaboylu_2014_td}(see Fig.~\ref{trajectories} A)). Thus, in the deep-tunneling regime the attoclock technique based on the photoelectron momentum measurement cannot reveal any deviation from the quasi-classical theory.

However, with increasing intensity the width of the tunneling barrier decreases and the time $\tau_W$ gets shorter such that it cannot overcome the effect of the initial momentum any more. In this so-called near-threshold tunneling regime the Wigner trajectory overtakes the simple-man one leading to an effectively negative asymptotic time delay~\cite{Yakaboylu_2013_rt, Yakaboylu_2014_td}(see Fig.~\ref{trajectories} B)). 

In order to embed the results of the quantum mechanical description of the tunneling step into the further classical propagation we define a classical trajectory starting at $t_i = \tau_W + t_s$ with an initial momentum $p_W$, which mimics the Wigner trajectory in the classical region (see the purple solid lines in Fig.~\ref{trajectories}). However, once we include the time $\tau_W$ into the formalism we have to take into account the field change during this time interval.  In the latter instant the TISE reads
\be 
\left( - \frac{\bm{\nabla}^2}{2} - \frac{1}{r} - \frac{\Phi(r)}{r} - \alpha_I \frac{\vec{r} \cdot \vec{E}(t_s + \tau_W)}{r^3}  + \vec{r} \cdot \vec{E}(t_s + \tau_W) \right ) \psi(\vec{r}) = -I_p \psi(\vec{r})\, .
\ee
After the following coordinate transformation
\be
x = \frac{x' \cos(\omega \tau_W) + \zeta \, y' \sin(\omega \tau_W)}{\sqrt{\cos^2(\omega \tau_W) +\zeta^2 \sin^2(\omega \tau_W)}} \, , \quad y = \frac{\zeta \, x' \sin(\omega \tau_W) - y' \cos(\omega \tau_W)}{\sqrt{\cos^2(\omega \tau_W) +\zeta^2 \sin^2(\omega \tau_W)}} \, , \quad z = z' \, ,
\ee
the TISE becomes
\be
\label{tise_1}
\left( - \frac{\vec{\nabla'}^2}{2} - \frac{1}{r'} - \frac{\Phi(r')}{r'} + \alpha_I \frac{x' \, \mathcal{G'}(t_s)}{r'^3}  - x' \, \mathcal{G'}(t_s) \right ) \psi(\vec{x'}) = -I_p \psi(\vec{x'}) \, ,
\ee
where $\vec{\nabla'} = \del / \del \vec{x'}$, $r' = \sqrt{x'^2 + y'^2 +z'^2}$, and $\mathcal{G'}(t_s) = \mathcal{G}(t_s + \tau_W) \sqrt{\cos^2(\omega \tau_W) +\zeta^2 \sin^2(\omega \tau_W)}$. As Eq.~(\ref{tise_1}) is in the form of Eq.~(\ref{tise_0}) we can separate the equation in parabolic coordinates and iterate the tunnel dynamics. However, since the applied ellipticity in the experiment is very close to the circular polarization, and $\mathcal{G}(t_s + \tau_W) \sim \mathcal{G}(\tau_W)$, the amplitude of the electric field remains almost the same $\mathcal{G'}(t_s) \sim \mathcal{G}(t_s)$. Hence, the Wigner time $\tau_W$ will be the same during the field rotation. Nevertheless the field rotates which affects the tunneling direction. As a consequence, the initial momentum and the tunnel exit in Cartesian coordinates result to
\bal
\label{mom_rot}
\vec{p}_i & \approx  \vec{p}_W \equiv p_W \frac{\cos(\omega \tau_W) \hat{\vec{x}} + \zeta \sin(\omega \tau_W) \hat{\vec{y}} }{\sqrt{\cos^2(\omega \tau_W) +\zeta^2 \sin^2(\omega \tau_W)}} \, , \\
\label{exit_rot}
\vec{r}_{exit} &\approx x_{exit} \frac{\cos(\omega \tau_W) \hat{\vec{x}} + \zeta \sin(\omega \tau_W) \hat{\vec{y}} }{\sqrt{\cos^2(\omega \tau_W) +\zeta^2 \sin^2(\omega \tau_W)}} \, .
\eal

In this improved final picture we reveal - in addition to the tunneling time and an initial longitudinal momentum - an initial transversal momentum as well as a correction for the tunnel exit.

\section{Comparison of Experiment and Theory}

\begin{figure}
  \centering
  \includegraphics[width=0.49\linewidth]{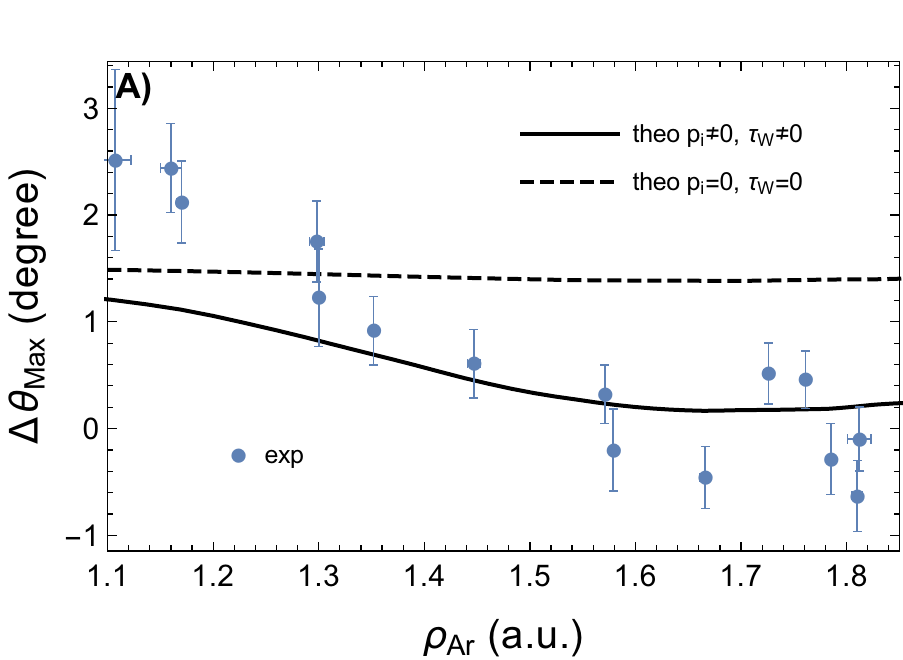}
  \includegraphics[width=0.49\linewidth]{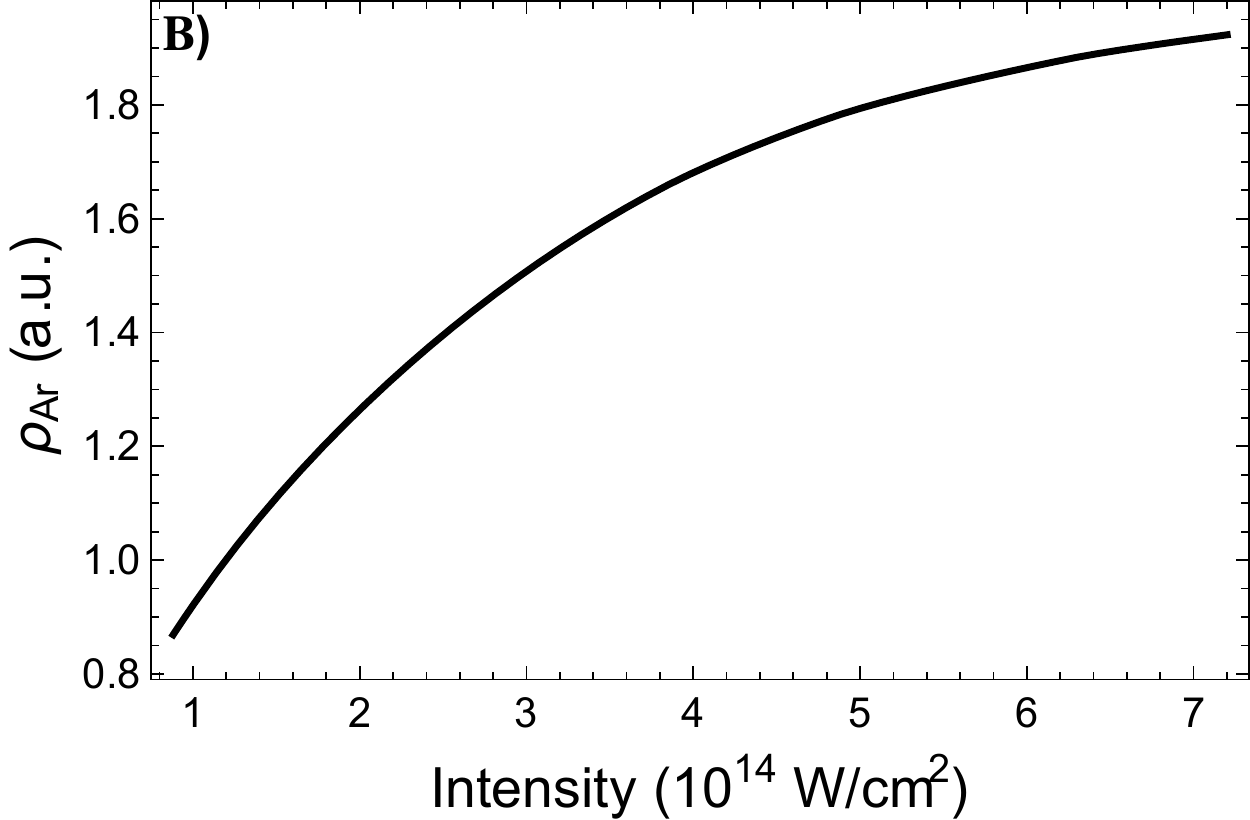}
  \caption{Figure A) shows comparison of the experiment and the theory for the angle difference. $\vec{p}_i$ is given by Eq.~(\ref{mom_rot}), and $\tau = \tau_W$. The applied parameters for the theory are $\zeta = 0.85$, $\tau = 36 \times 2 \pi / \omega= 55 \, \mbox{fs}$ with the envelope $cos^4 (\pi t/\tau)$. Figure B) demonstrates the theoretical prediction of the radius of the electron momentum distribution of Ar as a function of intensity.}
  \label{ang}
\end{figure}

After including all these effects the comparison of theory with experiment is shown in Fig.~\ref{ang} A) (see also main text). We note that the only differences between the two theoretical curves in Fig.~\ref{ang} A) are the chosen combinations for the initial momentum and the Wigner time, otherwise all the mentioned effects are included. In figure \ref{ang} B) the relation between the average electron momentum $ \rho_{Ar}$(the radius of the electron distribution for argon) and the applied intensity is shown. From Fig. \ref{ang} we conclude that the simple-man model (the dashed curve) fails to predict the experimental data whereas the solid curve, whose initial conditions are given by the Wigner formalism, agrees well with the experiment in the regime where $\gamma \ll 1$ or equivalently $ \rho_{Ar} \ge 1.3 a.u.$. In the low intensity domain ($ \rho_{Ar} < 1.3 a.u. $), where the Wigner formalism converges to the simple-man model because the Keldysh parameter approaches unity, one has to take into account non-adiabatic effects to tunnel-ionization such as non-adiabatic momentum shift~\cite{Mur_2001}, and tunnel exit shift~\cite{Klaiber_2014}.

\begin{figure}
  \centering
  \includegraphics[width=0.49\linewidth]{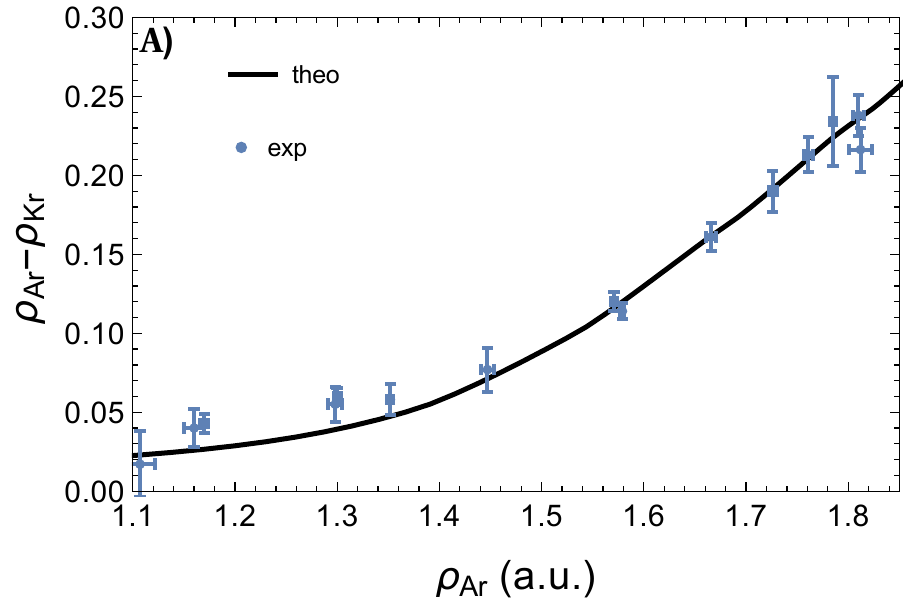}
    \includegraphics[width=0.49\linewidth]{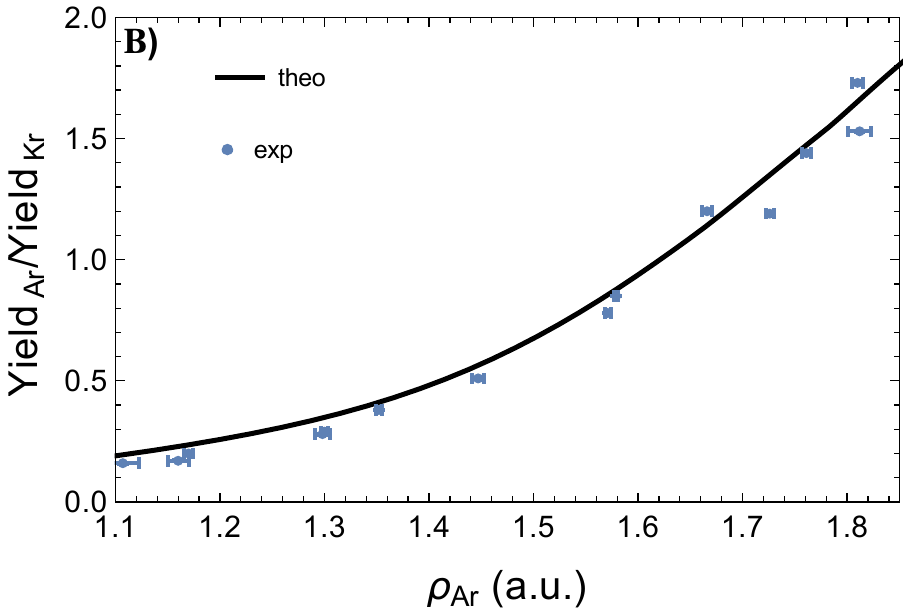}
  \caption{Comparison of the experiment and the theory; figure A) the difference between the average electron momentum, figure B) the yield ratio.}
  \label{rad_yield}
\end{figure}

As a final point, we emphasize that the pulse duration, shape and ellipticity are determined in the experiment with a limited precision. However, these parameters influence the outcome of the experiment. In order to put further constraints on their absolute values, we compare the theoretical results with two other observables: the difference between the average electron momentum $ \rho_{Ar}-\rho_{Kr}$ and the yield ratio between argon and krypton as a function of intensity. The latter can be calculated via
\be
\label{focal_yield}
\bar{Y} (I_0) = \frac{\int_0^{I_0} d I\, F (I,I_0) Y(I)}{\int_0^{I_0} d I\, F (I,I_0) } \, .
\ee
The results are shown in Fig.~\ref{rad_yield} A) and B), respectively. 

\begin{figure}
  \centering
\includegraphics[width=0.49\linewidth]{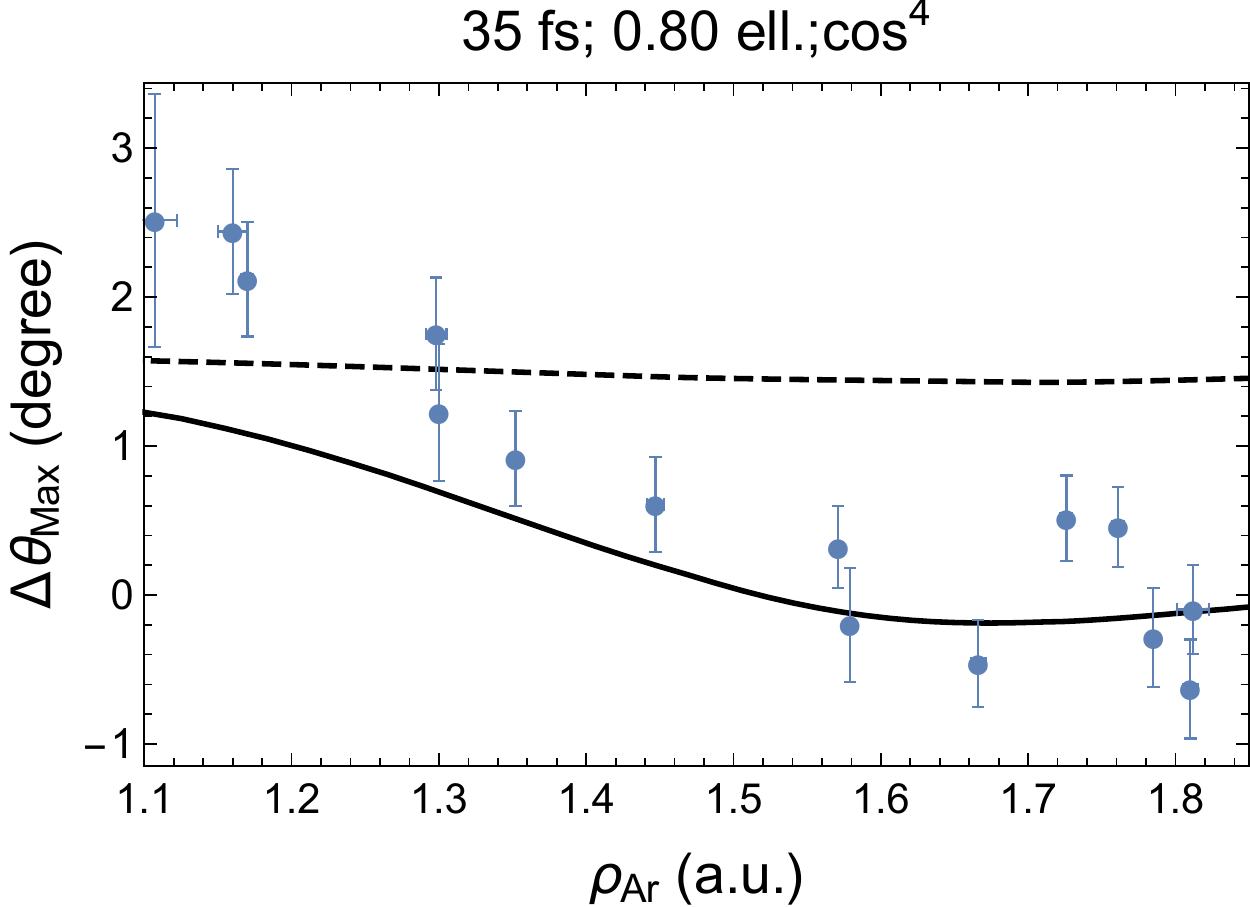}
 \vspace{0.5cm}
 \includegraphics[width=0.49\linewidth]{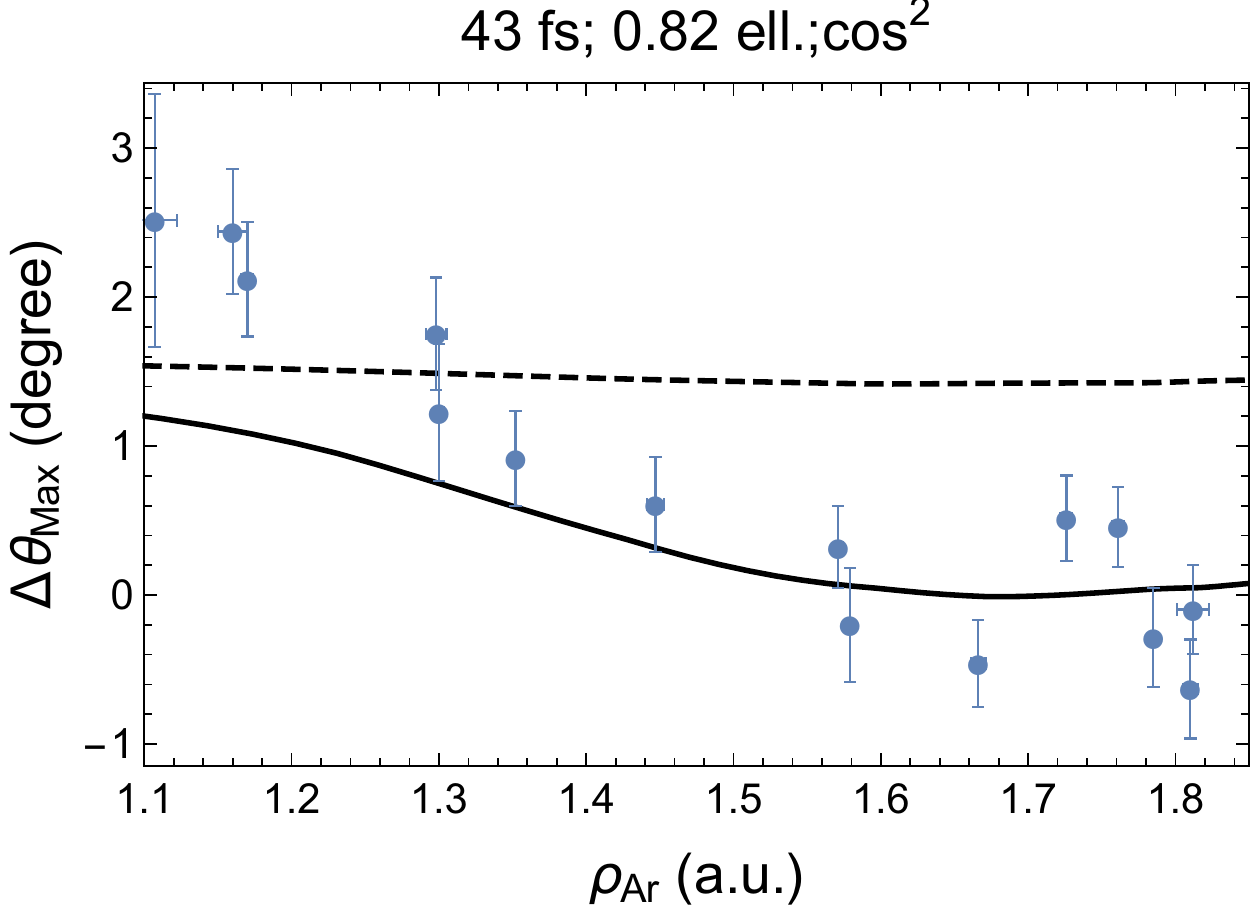} 
     \vspace{0.5cm}
  \includegraphics[width=0.49\linewidth]{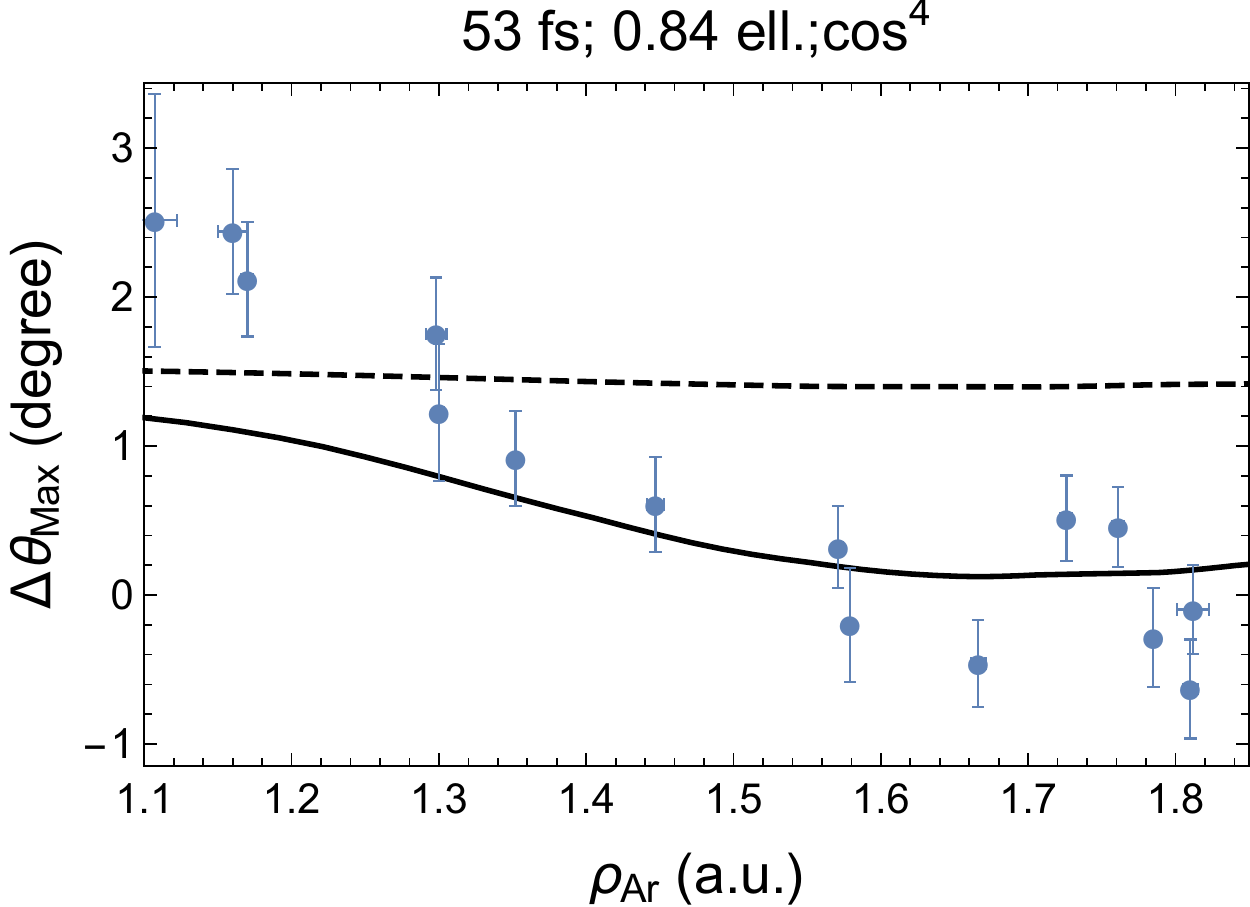}
   \includegraphics[width=0.49\linewidth]{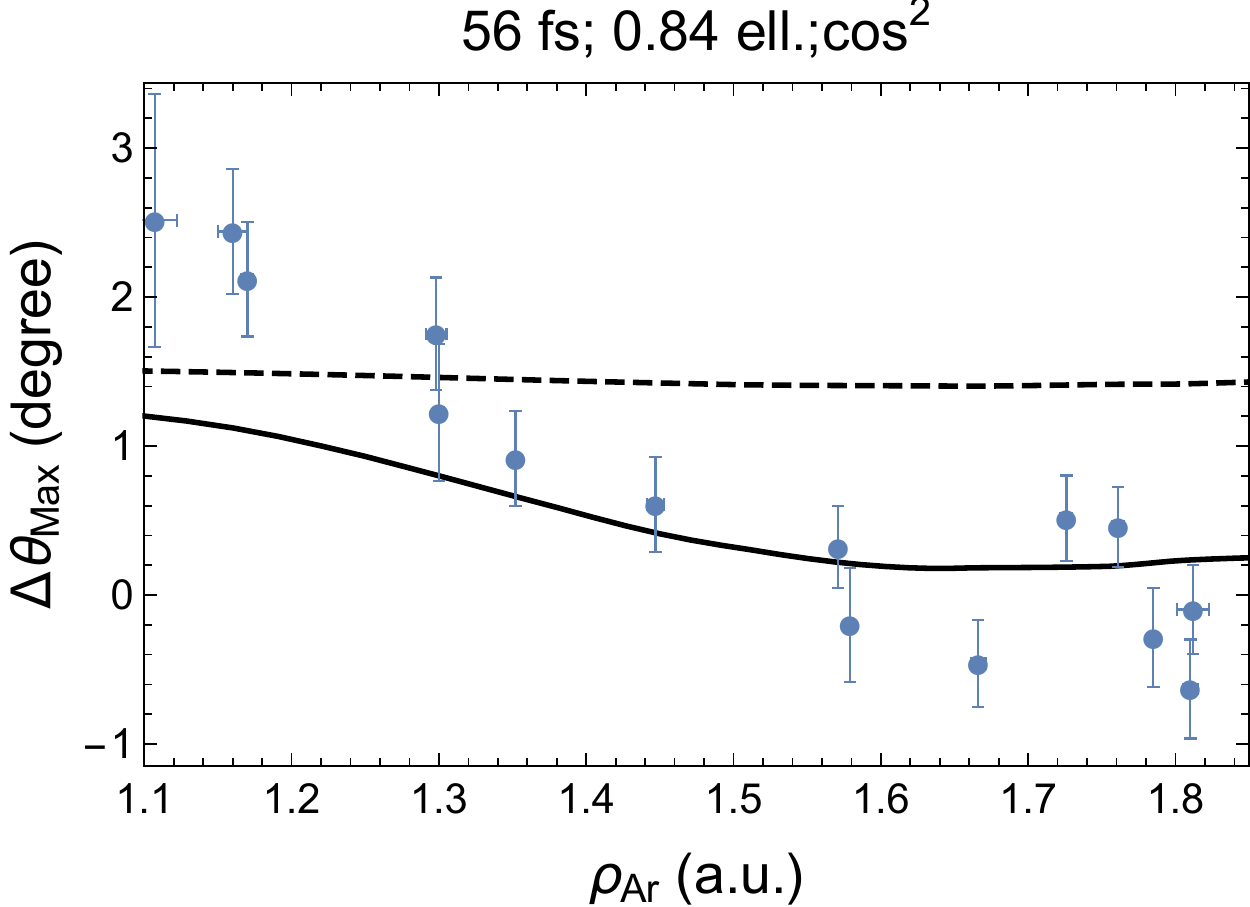}
    \includegraphics[width=0.49\linewidth]{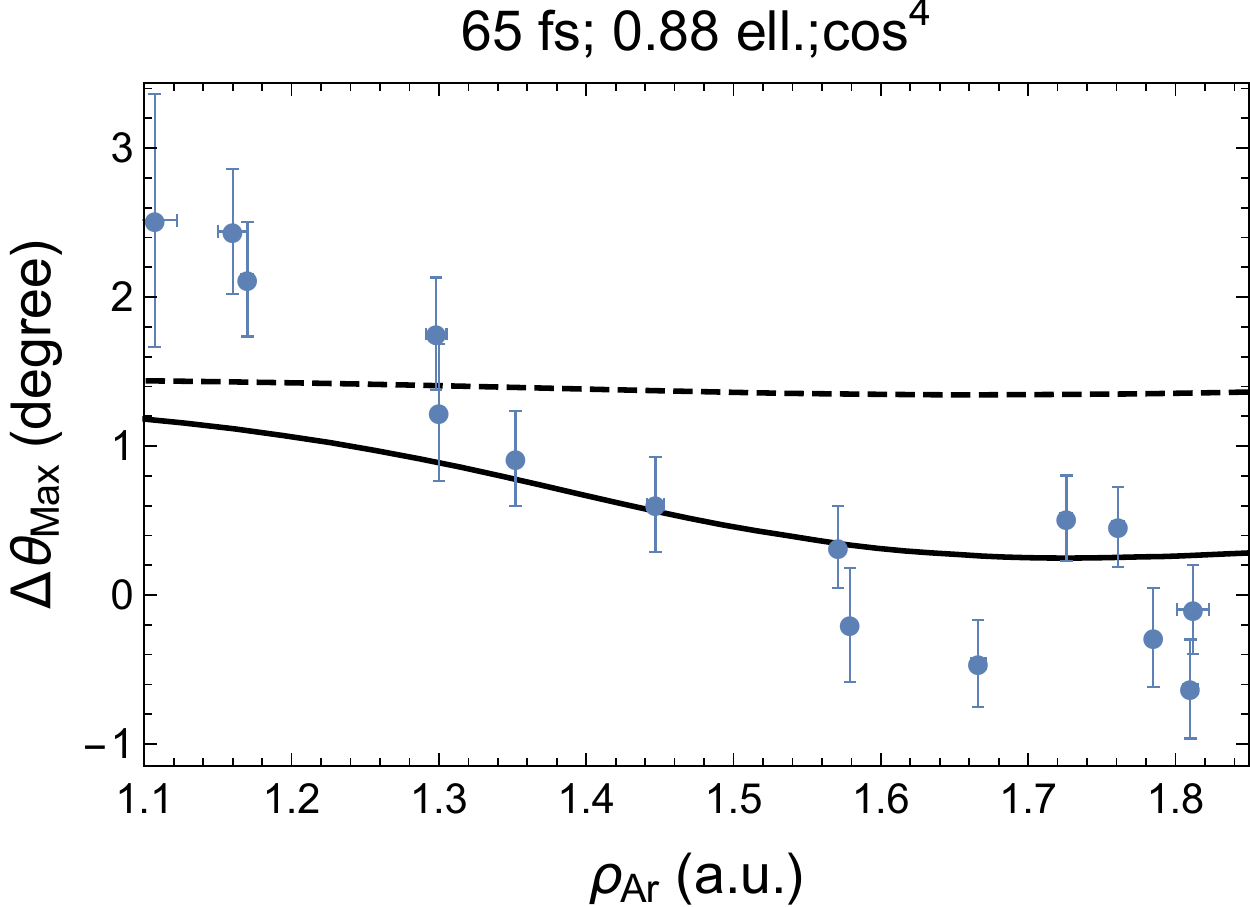}
     \includegraphics[width=0.49\linewidth]{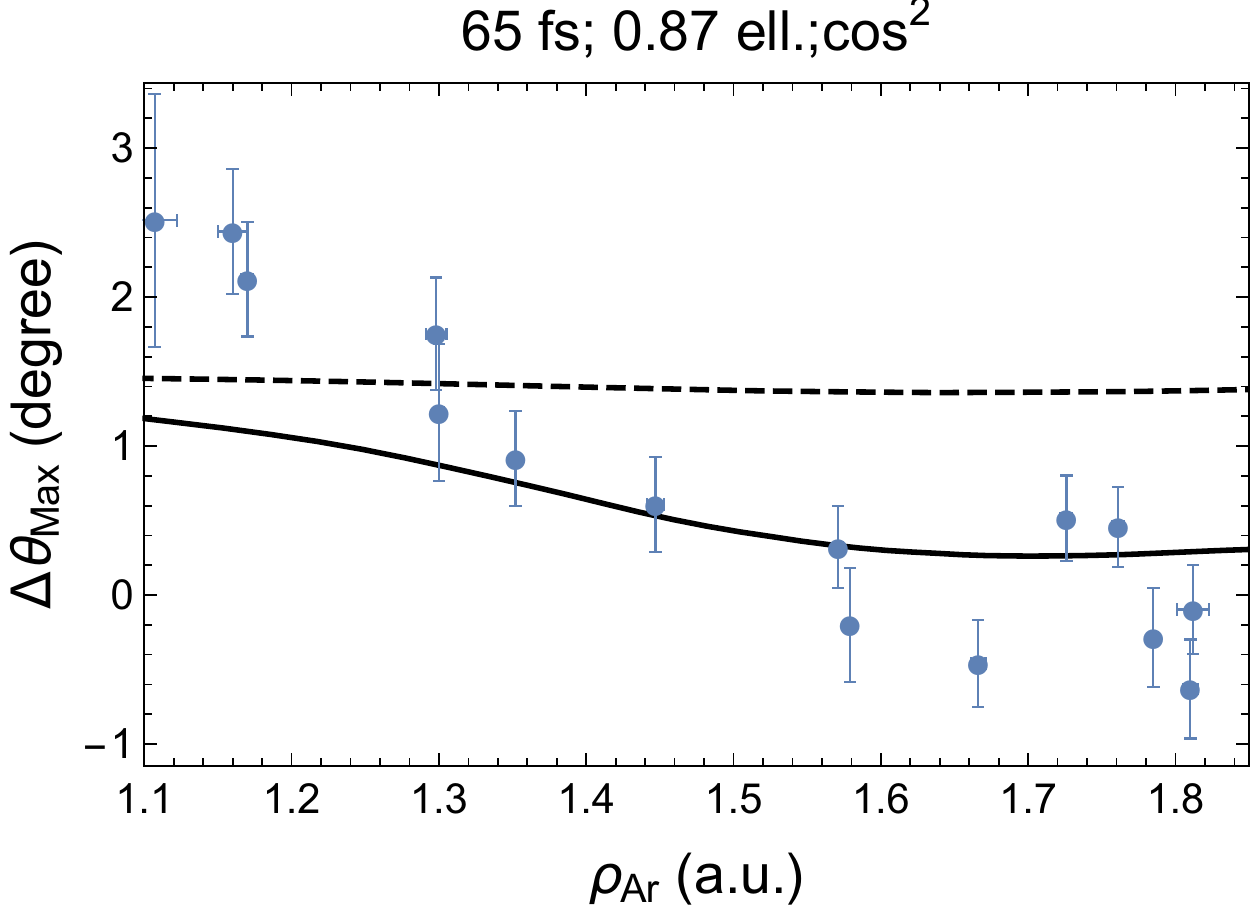} 
  \caption{Comparison of the experiment and the theory for different pulse durations, ellipticity, and envelopes.}
  \label{diff_ell}
\end{figure}

The free parameters describing the pulses (pulse shape, pulse duration, ellipticity) are varied over a range larger than the experimental uncertainties and for each set of parameters the intensity of the pulses is chosen such that these two additional observables are fitted by the theoretical curves. For each set of parameters, with the intensity calibrated, we then obtain - similar than for Fig.~\ref{ang} A) - the angle differences $\Delta(\theta_{max})$ in the theoretical curves and compare it with the experiment. Such curves are shown in Fig.~\ref{diff_ell} for a few set of parameters. We conclude that the variation of pulses parameters has almost no influence on the main observable $\Delta(\theta_{max})$. In order to convincingly explain the angle differences it is necessary to change the initial conditions of the tunneling electron according to the Wigner formalism.

\FloatBarrier

\clearpage 

\begin{thebibliography}{10}
\bibitem{Simpleman2}
\bibinfo{howpublished}{This approach named as a simple-man theory, has been
  proposed about 30 years ago in different variations by several authors and
  proved to have a high predictive power in discussions of strong-field
  ionization phenomena.}
\end{thebibliography}


\end{document}